\documentclass[letterpaper,11pt]{article}
\def\thefootnote{\arabic{footnote}}

\def\mG{m_{3/2}}
\def\m{\bar m}
\def\s{\bar s}
\def\S{\bar S}
\def\T{\bar T}
\def\t{\bar t}
\def\half{{1\over2}}
\def\tc{\tilde{c}}
\def\talph{\tilde{\alpha}}
\def\vev{$vev$}

\DeclareMathAlphabet   {\mathsc}{OT1}{cmr}{m}{sc}

\def\bea{\begin{eqnarray}}
\def\eea{\end{eqnarray}}
\def\beq{\begin{equation}}
\def\eeq{\end{equation}}

\def\[{\left [}
\def\]{\right ]}
\def\({\left (}
\def\){\right )}
\newcommand{\lang}{\left\langle}
\newcommand{\rang}{\right\rangle}

\newcommand{\lbr}{\left\{}
\newcommand{\rbr}{\right\}}
\newcommand{\oline}[1]{\overline{#1}}
\newcommand{\jbar}{\bar{\jmath}}
\newcommand{\ibar}{\bar{\imath}}

\newcommand{\wtd}[1]{\widetilde{#1}}
\newcommand{\Lag}{\mathcal{L}}
\newcommand{\D}{\mathcal{D}}
\newcommand{\wh}[1]{\widehat{#1}}
\newcommand{\h}[1]{\hat{#1}}

\newcommand{\notD}{\not{\hspace{-.05in}D}}

\newcommand{\GeV}      {~\mathrm{GeV}}

\newcommand{\GS}       {\mathsc{gs}}
\newcommand{\PL}       {\mathsc{pl}}
\newcommand{\PV}       {\mathsc{pv}}

\newcommand{\STR}      {\mathsc{str}}

\newcommand{\QUAD}     {\mathsc{quad}}

\newcommand{\superint}{\int\diff^{4}\theta}

\newcommand{\hc}       {\mathrm{\; h.c. \;}}

\newcommand{\pp}{\partial}

\newcommand{\STr}{{\rm STr}}
\newcommand{\Tr}{{\rm Tr}}

\newcommand{\diff}{\mbox{d}}
\newcommand{\order}{\mathcal{O}}
\newcommand{\re}{{\rm Re}}

\newcommand{\lra}{\leftrightarrow}
\newcommand{\loopp}{{1\over32\pi^2}}
\newcommand{\gappeq}{\mathrel{\rlap {\raise.5ex\hbox{$>$}}
{\lower.5ex\hbox{$\sim$}}}}
\newcommand{\lappeq}{\mathrel{\rlap{\raise.5ex\hbox{$<$}}
{\lower.5ex\hbox{$\sim$}}}}

\begin{document}
\begin{titlepage}
\begin{center}
            \hfill    LBNL-59069\\
            \hfill    UCB-PTH-05/38\\
            \hfill    UPR-1136-T \\
\end{center}

\begin{center}
{\large \sc On Quadratic Divergences in Supergravity, \\ Vacuum
Energy and the Supersymmetric Flavor Problem}
\end{center}

\bigskip

\begin{center}
Mary~K.~Gaillard$^{1}$ and Brent~D.~Nelson$^{2}$
\end{center}

\begin{center}
$^{1}${\it Department of Physics, University of California \\

and \\

Theoretical Physics Group, Lawrence Berkeley National Laboratory,
Berkeley, CA 94720, USA}
\end{center}

\begin{center}
$^{2}${\it Department of Physics and Astronomy,\\

University of Pennsylvania, Philadelphia, PA 19104, USA}
\end{center}

\begin{center} Dated: November 18, 2005 \end{center}

\begin{quotation} We examine the phenomenological consequences of
quadratically divergent contributions to the scalar potential in
supergravity effective Lagrangians. We focus specifically on the
effect of these corrections on the vacuum configuration of scalar
fields in softly-broken supersymmetric theories and the role these
corrections play in generating non-diagonal soft scalar masses.
Both effects can only be properly studied when the divergences are
regulated in a manifestly supersymmetric manner -- something which
has thus far been neglected in past treatments. We show how a
supersymmetric regularization can impact past conclusions about
both types of phenomena and discuss what types of high-energy
theories are likely to be safe from unwanted flavor-changing
neutral current interactions in the context of supergravity
theories derived from heterotic string compactifications.
\end{quotation}
\end{titlepage}

\newpage
\renewcommand{\thepage}{\roman{page}}
\setcounter{page}{2}
\mbox{ }

\vskip 1in

\begin{center}
{\bf Disclaimer}
\end{center}

\vskip .2in

\begin{scriptsize}
\begin{quotation}
This document was prepared as an account of work sponsored by the United
States Government. Neither the United States Government nor any agency
thereof, nor The Regents of the University of California, nor any of their
employees, makes any warranty, express or implied, or assumes any legal
liability or responsibility for the accuracy, completeness, or usefulness
of any information, apparatus, product, or process disclosed, or represents
that its use would not infringe privately owned rights. Reference herein
to any specific commercial products process, or service by its trade name,
trademark, manufacturer, or otherwise, does not necessarily constitute or
imply its endorsement, recommendation, or favoring by the United States
Government or any agency thereof, or The Regents of the University of
California. The views and opinions of authors expressed herein do not
necessarily state or reflect those of the United States Government or any
agency thereof of The Regents of the University of California and shall
not be used for advertising or product endorsement purposes.
\end{quotation}
\end{scriptsize}

\vskip 2in

\begin{center}
\begin{small}
{\it Lawrence Berkeley Laboratory is an equal opportunity employer.}
\end{small}
\end{center}

\newpage

\renewcommand{\thepage}{\arabic{page}}
\setcounter{page}{1}
\def\thefootnote{\arabic{footnote}}
\setcounter{footnote}{0}


\section*{Introduction}
The supersymmetric flavor
problem~\cite{Ellis:1981ts,Donoghue:1983mx} continues to lie at
the heart of phenomenological treatments of supersymmetry
breaking. The apparently small size of off-diagonal mass terms in
the effective potential in any softly broken supersymmetric model
has often been cited as particularly troublesome for models in
which supergravity plays a prominent role in transmission of
supersymmetry breaking to the observable sector -- and therefore
for string-inspired models more generally~\cite{Louis:1994ht}.

It is important, however, to distinguish ``general gravity
mediation'' from the specific manifestations that arise in string
models. In a completely general model of gravity mediation there
is indeed a generic ``flavor problem,'' irrespective of the
quadratic divergence issue. That is to say that one can imagine
{\em a priori} operators of the form
\begin{equation}
\superint \frac{\oline{X} X}{M_{\PL}^{2}} \oline{Q}^{\bar{i}} Q^j,
\label{bad} \end{equation}
where $X$ is a Standard Model singlet that is presumably a hidden
sector field. If it participates in supersymmetry (SUSY) breaking,
then it will generate off-diagonal soft-masses. The key is that
without specifying a rule for how this $X$ couples to Standard
Model matter (such as via gauge charges in gauge mediation) then
we must assume that different flavors can be treated differently.
Rephrased in a manner common to treatments of the flavor problem
in supersymmetry: there is no symmetry argument as to why
operators of the form of~(\ref{bad}) should be absent.

But the symmetry here being considered is typically that of a {\em
gauge symmetry}. Yet the operator in~(\ref{bad}) may admit a {\em
geometrical} interpretation. Let us rewrite things to make this
more apparent
\begin{equation}
\superint \frac{R_{\bar{i}j\bar{k}\ell}}{M_{\PL}^{2}}
\oline{X}^{\bar{k}} X^{\ell} \oline{Q}^{\bar{i}} Q^j ,
\label{bad2} \end{equation}
where the tensor $R_{\bar{i}j\bar{k}\ell}$ is the curvature tensor
formed from the field-reparameterization connection
\begin{equation}
R^{i}_{jk\bar{m}} = D_{\bar{m}} \Gamma^{i}_{jk}; \quad
\Gamma^{i}_{jk} =  K^{i\bar{n}}\partial_{j} K_{k\bar{n}} =
K^{i\bar{n}}\partial_{j}\partial_{k}\partial_{\bar{n}}K,
\label{gammas} \end{equation}
where $D_{\bar{m}} = K_{\ell\bar{m}}D^{\ell}$ is a covariant
derivative with respect to field reparameterization and
$K_{\ell\bar{m}}$ is the K\"ahler metric. Thus, while an
understanding of the form of this tensor in~(\ref{bad2}) may not
be possible in terms of the gauge quantum numbers of the fields
involved, an understanding in terms of the isometries of the
manifold defined by the chiral superfields of the theory may
indeed exist. This point has been emphasized recently for
string-based effective supergravity
theories~\cite{Chankowski:2005jh,Lebedev:2005uh}.

In the absence of an underlying theory we typically proceed by
making assumptions. For example, we can achieve the minimal
supergravity paradigm by assuming the curvature tensor factorizes
as $R_{\bar{i}j\bar{k}\ell} \propto K_{\bar{i}j}K_{\bar{k}\ell}$
and $K_{\bar{i}j} \propto
\delta_{\bar{i}j}$~\cite{Barbieri:1982eh,Chamseddine:1982jx}. In
string-based models there may be a basis in which the K\"ahler
metric for the matter fields is diagonal in its generation
indices~\cite{Binetruy:2005ez}. This need not be the same basis in
which fermion masses are diagonal, but this is the standard
problem that all methods of SUSY breaking share. Then the question
is whether higher order terms such as~(\ref{bad2}) are the result
of terms at higher genus in the string loop expansion (and thus
require a string-theory explanation) or are the result of
field-theory loop corrections involving gravitational-strength
interactions. In the case of the latter, the all-important matrix
$R_{\bar{i}j\bar{k}\ell}$ can be determined from the tree-level
theory.

But the supergravity flavor problem is often considered even more
severe than the discussion above would indicate. As a
non-renormalizable effective theory, supergravity models suffer
corrections to their effective potentials that grow quadratically
with some high mass scale. The structure of these loop-induced
contributions to the effective potential is determined by the
structure of the tree-level K\"ahler potential for matter fields.
While these corrections come with a loop-factor relative to the
leading-order contributions, it has been noted that they are often
also proportional to the number of light fields in the
theory~\cite{Choi:1994xg,Choi:1997de}, which can be quite large
for string-based models. Thus, it is argued, it may not be
sufficient to ensure the absence of dangerous terms such
as~(\ref{bad2}) at the leading order. As these terms also
contribute to the vacuum expectation value of the effective
potential, their presence has been conjectured to play a
significant role in determining the vacuum configuration of light
fields in the theory.

In this paper we re-visit these issues, taking care to address
them only after a proper (supersymmetric) regularization of the
apparent divergences is performed. This important step has been
completely overlooked in previous treatments of the
phenomenological implications of quadratic divergences. In
Section~\ref{sec:general} below we exhibit the divergent
contributions to the effective potential in a non-technical manner
and discuss the importance of a proper regularization scheme. We
intend for the treatment in that section to be accessible to the
non-expert. In Section~\ref{sec:technical} we present the
technical details, first of the quadratic divergences themselves
in Section~\ref{sec:quad} and then in Section~\ref{sec:PV} of the
Pauli-Villars (PV) regulator sector we will use to cancel the
divergences and render these contributions finite. In
Section~\ref{sec:renorm} we demonstrate explicitly how the
supersymmetric regularization scheme allows for an interpretation
of the loop corrections as a supersymmetric renormalization of the
space-time metric and K\"ahler potential themselves. The reader
uninterested in this level of technical detail can proceed
directly to Section~\ref{sec:phenom}, where the phenomenological
implications of these loop corrections for vacuum stabilization
and scalar masses are treated. This section is largely
self-contained, with only the occasional need to refer to certain
key results from Section~\ref{sec:technical}.

\section{Divergent Contributions at One Loop to the Effective Scalar
Potential}
\label{sec:general} The effective potential at one loop for a generic theory is given
by~\cite{Coleman:1973jx}~-~\cite{Chiou-Lahanas:1992am}
\begin{equation}
V_{\rm 1-loop} = V_0 + \frac{1}{64\pi^2}\STr \mathcal{M}^0 \cdot
\Lambda^4 \ln \frac{\Lambda^2}{\mu^2} + \frac{1}{32\pi^2}\STr
\mathcal{M}^2 \cdot \Lambda^2 + \frac{1}{64\pi^2} \STr
\mathcal{M}^4 \cdot \ln \frac{\mathcal{M}^2}{\Lambda^2} + {\rm
finite} ,
\label{Vtotal} \end{equation}
where we define the quantity
\begin{equation}
\STr \mathcal{M}^{n} \equiv \sum_i (-1)^{2 J_i} C(J_i) \; m_i^n ,
\label{STrM} \end{equation}
with $m_i$ being the (possibly field-dependent) mass of the state
in the summation, and $C(J_i) = 2J_i +1$ for $J_i \leq 1$. The
definition of the supertrace over the degrees of freedom in the
gravity sector depends on the gauge-fixing procedure
utilized~\cite{Gaillard:1993es}.\footnote{For example
$C(\frac{3}{2}) = + 4$ in the gauge
of~\cite{Gaillard:1993es,Gaillard:1996hs,Gaillard:1996ms} that is
used in the results quoted below.} The first term $V_0$ is the
tree-level (classical) scalar potential. The second quantity is,
strictly speaking, the loop-induced ``vacuum energy'' in the
limited sense that it is a contribution to the scalar potential
that scales like the fourth power of the cut-off $\Lambda$ with a
field-independent coefficient. In any model where the spectrum
obeys $N=1$ supersymmetry this coefficient vanishes identically.
The third and fourth terms in~(\ref{Vtotal}) are the quadratic and
logarithmic divergences, respectively. These were computed for a
general supergravity Lagrangian
in~\cite{Srednicki:1984un}~-~\cite{Gaillard:1996ms} and for
specific heterotic string models with modular invariance
in~\cite{Binetruy:1985ap,Binetruy:1987cn,Burton:1989ai}.

Before proceeding to the issue of scalar mass and flavor-changing
effects in the low-energy theory, let us pause for a brief
digression on the issue of vacuum energy in these theories. Given
that we expect the masses of heavy fields in the theory to be on
the order of the gravitino mass such that $m_i \sim m_{3/2}$, we
might imagine that the quadratically divergent term
in~(\ref{Vtotal}) is providing a contribution to the vacuum energy
in an amount $\lang \delta V_{\rm 1-loop} \rang \simeq m_{3/2}^2
M_{\PL}^2$. For consistency with cosmological observations (and
for the sake of doing a meaningful phenomenological study of
effective Lagrangians) we are often at pains to engineer the
vanishing of the vacuum expectation value (\vev) of the tree-level
scalar potential: $\lang V_0 \rang =0$. A straightforward
interpretation of~(\ref{Vtotal}) would seem to suggest that the
one-loop contribution to this effective vacuum energy is generally
positive and that we should therefore strive to engineer $\lang
V_{0} \rang <0$. This was the standpoint of~\cite{Choi:1994xg},
where it was suggested that this contribution to the vacuum energy
might compensate for the large negative contribution that
typically arises in the so-called ``racetrack'' models of moduli
stabilization.

It is certainly true that whatever mechanism Nature employs in
ensuring a vanishing (or nearly vanishing) cosmological constant
must do so to amazing accuracy; surely a phenomenological model of
this mechanism (such as the minimization of some potential for a
modulus which couples to these various contributions) must take
into account these loop-induced terms. The presence of the loop
factors in~(\ref{Vtotal}) might suggest that these higher-order
contributions are small perturbations on the result that causes
the classical vacuum energy to vanish. But, as was pointed out
in~\cite{Choi:1994xg}, the sum over multiplets implied in the
supertrace brings a potentially large number into the one-loop
contribution. For example, in a simple model with only $N_{\chi}$
chiral superfields, $N_G$ Yang-Mills multiplets, and gravity we
would have~\cite{Gaillard:1993es,Gaillard:1996ms}
\begin{equation}
\lang \delta V_{\rm 1-loop} \rang \sim \frac{\Lambda^2}{32\pi^2}
\STr \mathcal{M}^2 \simeq \frac{\Lambda^2}{16\pi^2}\(N_{\chi}
m_0^2 -N_G m_{1/2}^2 + 2 m_{3/2}^2\)
\label{STr} \end{equation}
with the coefficient of the last term in~(\ref{STr}) depending on
the gauge-fixing prescription employed. For the MSSM field content
(without right-handed neutrinos) these coefficients are $N_{\chi}
= 49$ and $N_G = 12$. For the much larger field content of a
typical $Z_3$ orbifold compactification of the $E_8 \times E_8$
heterotic string we might have 300 or more chiral multiplets while
the gauge multiplets are typically bounded with $N_G \lappeq
65$~\cite{Giedt:2001zw}.

But the conclusion that this contribution is then typically
positive is a naive one. In particular, the form of~(\ref{STr})
implies a common cut-off for each term, whereas these cut-offs
will vary from term to term and may be field-dependent. Indeed,
only by employing field-dependent cut-offs can supersymmetry be
maintained in the presence of loop corrections to the effective
potential~\cite{Binetruy:1988dw}~-~\cite{Gaillard:1994sf}. When
the effective cut-off is field-dependent it is no longer correct
to assume that $\Lambda_{\rm eff}^{2} > 0$ in all cases. In fact,
even if the effective cut-offs are constants, four-dimensional
supergravity theories require {\it a priori} at least two
subtractions to render the divergent loop integrals finite. In the
context of Pauli-Villars regularization this is equivalent to
requiring at least two sets of fields for every one light field.
Proper regularization of the theory -- which is to say,
elimination of divergences -- thus results in the replacement
\begin{equation}
\Lambda^2 \STr\mathcal{M}^2 \to \STr \mu^2 \mathcal{M}^2 \ln(\mu^2)
\eta_S, \quad \quad \eta_S = \sum_q \eta_q \lambda_q \ln \lambda_q .
\label{subtract} \end{equation}
The quantity $\eta_S$ is a coefficient that represents the
uncertainty in the threshold for the onset of this new physics and
$\eta_q = \pm 1$ is chosen to ensure finiteness of the effective
Lagrangian. Note that we can define the quantity $\mu^2_q =
\lambda_q \mu^2$, $\lambda_q >0$ which represents the mass-scale of
the new physics corresponding to each of the $q$ subtractions (or PV
superfields $\Phi^q$) that are serving as regulators to the
supergravity theory.   For the case of regularization via
Pauli-Villars (PV) regulators, which we will employ in the next
section, these threshold factors can be computed in terms of the
properties of the PV sector. The sign of the effective cut-off is
determined by the sign of the factor $\eta_S$.

It turns out that the cancellation of all ultraviolet (UV)
divergences in realistic string-derived supergravity models
requires at least~5 PV supermultiplets for each supermultiplet of
the low energy theory~\cite{bgpr}. The important point is that
with four or more subtractions it is not necessary for the
coefficient $\eta_S$ to be positive.\footnote{See, for example,
the discussion of this point in Appendix~C
of~\cite{Binetruy:1988nx}.} As this is an important point, let us
take a moment to sketch its derivation. The cancellation of
quadratic divergences imposes the following
conditions~\cite{Binetruy:1988nx} on the sum over signatures
\begin{equation}
\sum_{q=1}^S\eta_q = - 1,\qquad \sum_q\eta_q\lambda_q = 0,
\label{PVreplace} \end{equation}
where $S\ge2$ is the total number of subtractions (PV
superfields). The first equality requires $S$ odd: $S = 2n + 1\ge
3$. In the simplest case where $S = 3$, $\eta_q = (-1,-1,1)$, and
$\lambda_q = (\lambda_1,\lambda_2,\lambda_1 + \lambda_2)$, we have
\begin{equation}
\eta_S = -\lambda_1\ln\lambda_1 - \lambda_2\ln\lambda_2 +
(\lambda_1 + \lambda_2)\ln(\lambda_1 + \lambda_2) >0 .
\label{S3} \end{equation}
However for $n\ge1,\;S\ge5$, $\eta_S$ is not positive definite. To
see this take the following example
\bea \eta_1,\cdots\eta_n &=& 1, \qquad \quad
\eta_{n+1},\cdots\eta_{2n+1} = -1, \nonumber \\
\lambda_1\cdots\lambda_n &=& \Lambda/n, \qquad
\lambda_{n+1}\cdots\lambda_{2n} = \lambda, \qquad \quad
\lambda_{2n+1} = \Lambda - n\lambda \; .\eea
Then
\beq\eta_S = \Lambda\ln(\Lambda/n) - n\lambda\ln\lambda - (\Lambda
- n\lambda)\ln(\Lambda - n\lambda) = \lambda
f(\Lambda/\lambda,n).\eeq
Now since
\beq \lim_{x\to\infty}f(x,n) = - x\ln n + n\ln x + O(x^{-1})\eeq
is negative for $n>1$, the function $\eta_S$ is positive definite
only for $n=1,\;S=3$.  Cancellation of all the ultraviolet
divergences of a general supergravity theory requires at least 5
PV chiral multiplets for every light chiral multiplet and even
more PV supermultiplets to regulate the gauge loops. Therefore one
cannot assume that the effective cut-offs are positive.

In general, using a straight cut-off regulator is equivalent to an
explicit breaking of supersymmetry and is thus inconsistent with
the requirement that supersymmetry be broken only spontaneously by
the $vev$s of fields in the low-energy theory. Rather, these
cut-offs should be regarded as field-dependent. Indeed, it is the
very field-dependence of these cut-offs that allows for the
possibility of a dynamical mechanism for cancelling the vacuum
energy to arbitrarily high loop-order. We will discuss an example
of this possibility, including the effect of the one-loop
correction in Section~\ref{sec:vac} below. Here we wish to briefly
describe the nature of this field dependence. Supersymmetric
Pauli-Villars regularization of matter and gauge loops\footnote{As
noted below full regularization of gravity loops requires the
introduction of massive PV (Abelian) vector multiplets as well.}
is implemented by introducing supersymmetric masses for PV fields
$\Phi^A, \; \Pi^A$ through superpotential couplings of the form
\beq W_{\PV} = \mu_{A B}(Z^i)\Phi^A\Pi^B,\label{pvpot} \eeq
where $\mu_{A B}$ is a holomorphic function of the light chiral
multiplets $Z^i$.  Then the effective squared cut-off is replaced
by the (signature-weighted) squared PV mass matrix:
\beq \Lambda^2_{\rm eff}\to (M^2_{\PV})^A_B = e^K K_\Phi^{A\bar
C}K_\Pi^{D\bar E} \bar\mu_{\bar C\bar E}(\bar Z)\mu_{B
D}(Z).\label{PVfullmass}\eeq
By convention we denote by $\Phi^A$ those fields that regulate
divergences arising from the gauge and superpotential couplings of
the light fields; their K\"ahler metric $K^\Phi_{A\bar B}$ is
determined by the effective supergravity theory of the light
fields.\footnote{We will use upper-case letters to collectively
denote the fields in the PV sector throughout this work.} The
K\"ahler metric for the fields $\Pi$, introduced to generate
supersymmetric PV masses, is much less constrained. For example if
we took
\beq\mu_{A B} = \mu\delta_{A B}, \qquad K^\Pi_{A\bar B} =
e^{-K}(K_\Phi^{-1})_{A\bar B}, \label{PVscenA} \eeq
with $\mu$ constant, the effective cut-off would indeed be constant.
This may seem like a very convenient choice from the point of view
of the effective field theory. Considerations of the anomaly
structure of the effective theory in string-based models may make
such constant effective cut-offs inconsistent. In the effective
field theory of heterotic orbifold
models~\cite{Antoniadis:1991fh,Dixon:1990pc}, for example, the
anomaly structure does not allow for constant
cut-offs~\cite{Gaillard:1992bt}. Since superpotential couplings
depend on moduli in a general string model, this is likely to be a
generic feature of effective supergravity models which seek to
describe their low-energy behavior. As we will see in
Section~\ref{sec:phenom}, this dependence is crucially important for
the low-energy phenomenology of supergravity effective theories at
the one-loop level. We will have more to say about the implications
of~(\ref{PVfullmass}) in the conclusion section of this work.

But there is a deeper reason why the cut-offs in~(\ref{Vtotal})
should be regarded as field-dependent and not straight cut-offs. In
addition to the substitution $\Lambda^2\to M^2_{\PV}$ in the cut off
theory, the presence of the superpotential~(\ref{pvpot}) induces
additional terms in the regulated theory such that the net result of
the quadratically divergent contributions is {\em a renormalization
of the space-time metric and the K\"ahler potential itself}. By
simply inserting a cut-off in the divergent integrals one would not
generate these additional terms that are needed for
supersymmetry~\cite{Binetruy:1988nx,Gaillard:1994sf}. Put
differently, the use of straight cut-offs breaks the local
supersymmetry of the supergravity Lagrangian and prevents us from
treating the renormalization at the superspace level. As will be
made explicit in Section~\ref{sec:renorm} below, the interpretation
of the one-loop quadratic divergences in terms of
renormalizations is valid only to lowest order in the loop expansion
parameter $\epsilon = \hbar/16\pi^2$.  If these terms have very
large coefficients, all the leading $(\epsilon\Lambda_{eff}^2)^n$
terms must be retained so as to maintain manifest supersymmetry; the
full correct result is easily inferred from the one-loop
calculation.

Finally, let us note that nowhere do we make the artificial
distinction between contributions to vacuum energy arising from
``hidden'' versus ``observable'' sector fields. Such a separation
of contributions serves no physical purpose and creates confusion:
any dynamical mechanism engineered in the low-energy theory to
produce vanishing vacuum energy at the tree or $n$-loop level
should take into account contributions to~(\ref{Vtotal}) arising
from {\em all} sectors of the theory, including effects from
symmetry breaking at energy scales lower than that of
supersymmetry breaking (i.e. electroweak symmetry breaking, the
QCD phase transition, etc.), even though these will likely have
negligible impact on the actual vacuum configuration of the fields
involved. It is, after all, the entirety of vacuum energy that is
constrained by the apparent flatness of the universe (or measured,
if one likes, by the recession rate of supernovae).

Having treated the issue of vacuum energy let us return to our
primary concern: corrections to the scalar potential that may
produce dangerous flavor-changing operators. Within the supertrace
$\STr \mathcal{M}^2$ of the quadratically divergent term
in~(\ref{Vtotal}) we find terms such as
\begin{equation}
\STr\mathcal{M}^2 \ni 2e^{-K}R_{i\bar{j}}\oline{A}^{i}A^{\bar{j}}
=2 R_{m\bar{n}i\bar{j}} K^{m\bar{n}}K^{i\bar{p}}K^{q\bar{j}}
e^{K}(K_{\bar{p}}\oline{W} +\oline{W}_{\bar{p}})(K_q W + W_q)
\label{danger} \end{equation}
where we have introduced the notation $A \equiv e^K W$ for later
convenience. Several potentially dangerous contributions to the
effective scalar potential are contained within~(\ref{danger}),
among them a term with the structure of~(\ref{bad2})
\begin{equation}
\Lag_{\rm 1-loop} \ni \frac{\Lambda^2}{16\pi^2}
K^{m\bar{n}}R_{m\bar{n}i\bar{j}}(e^K|W|^2)Q^{i}\oline{Q}^{\bar{j}}
\label{bad3} \end{equation}
where the cut-off scale $\Lambda$ is understood to be measured in
units of the reduced Planck mass, which we have set to unity.
While this cut-off is presumably $\order(1)$ in these units, the
loop factor is partially compensated by a potentially large number
coming from the contraction on the first two indices in the
curvature tensor in~(\ref{bad3}). This term in the one-loop
effective Lagrangian has a form capable of producing an
off-diagonal scalar soft mass term for the squarks, which may be
of the same general size as the tree-level (and presumably
diagonal) soft masses~\cite{Choi:1997de,Choi:1997cm}. In the most
general supergravity theory we have no reason to expect that all
elements of the curvature tensor are not populated. The question
of whether or not such dangerous terms really exist in the low
energy theory then becomes a question of the properties of the
non-linear sigma model inherited from the underlying theory.

\section{Quadratic Divergences and Their Regulation}
\label{sec:technical}

Having established the crucial importance of the quadratically
divergent contributions to~(\ref{Vtotal}) in a schematic way in
the previous section, we now wish to exhibit the full structure of
these terms in a properly regularized context. In order to do this
we will need to introduce the Pauli-Villars sector in a more
complete manner. This section provides the necessary technical
details to understand the notation and origin of the results we
present in Section~\ref{sec:phenom}. This is not a complete
description of the PV technique in supergravity theories. We will
make certain simplifications (which we point out) along the way
that are made possible by our desire to study the quadratic
divergences only -- a full treatment of the logarithmic as well as
the quadratic divergences can be found
in~\cite{Gaillard:1996hs,Gaillard:1996ms}.

Regularization of supergravity theories is no simple matter. The
work that we attempt to summarize here spans over a decade of
research. Even the regularization of a simple system -- such as a
non-linear sigma model without Yang-Mills fields in a curved
background -- requires several species of PV regulator fields. The
proliferation of fields, both in number and type, grows rapidly
when we come to consider the string-inspired models in which we
are most interested (such as those with non-canonical gauge
kinetic terms in which Yang-Mills fields couple to a dilaton
field). This is an unfortunate yet unavoidable fact of working
with supergravity effective theories. To guide the reader we first
make a point about notation. In general, upper-case indices refer
to fields in the PV sector, while lower-case indices refer to
light fields of the theory. Fields such as the $\Phi^P$ introduced
in~(\ref{pvpot}) from the previous section are meant to represent
all fields of a certain type: here all the chiral superfields
$\Phi$ of the PV sector which regulate loops involving light
fields $Z$. Within this set we might refer to the fields $\Phi^I$
as that subset that transform under the gauge group(s) in the same
way as the field $Z^i$, or the fields $\Phi^a$ that transform as
the adjoint under the particular group $\mathcal{G}_a$. To each of
these subsets (or species) we occasionally must associate multiple
copies, labeled by Greek indices $\alpha$, $\beta$, etc. These are
the multiple copies that mimic the multiple insertions needed to
regulate the theory, as mentioned in Section~\ref{sec:general}. We
often suppress this index when it would be superfluous in a
particular expression, but it should be understood to be present.

\subsection{The Quadratically Divergent Contribution at One Loop}
\label{sec:quad} Our starting point is the one-loop effective action, which can be
determined from those terms quadratic in the quantum fields when the
Lagrangian is expanded about an arbitrary background
\begin{equation}
\Lag_Q = -\frac{1}{2} \Phi^{T} Z_{\Phi} (\hat{D}_{\Phi}^{2}
+H_{\Phi}) \Phi + \frac{1}{2} \oline{\Theta} Z_{\Theta}
(i\notD_{\Theta} - M_{\Theta}) \Theta + \Lag_{\rm gh} + \Lag_{\rm
Gh} ,
\end{equation}
where the last two terms represent ghost and ghostino Lagrangians,
respectively. The quantities $\Phi$ and $\Theta$ are column
vectors which contain the quantum bosons and quantum fermions,
respectively, of the theory. For example, $\Phi = (h_{\mu\nu},\;
\wh{A}^{a},\; \hat{Z}^{i}, \; \hat{\bar{Z}}^{\bar{m}})$ contains
the graviton, gauge and scalar quantum fields and is a $2N_{\chi}
+ 4N_{G} + 10$ component object. Similarly, with the gauge-fixing
choice of~\cite{Gaillard:1993es}~-~\cite{Gaillard:1996ms},
$\Theta$ is an $N_\chi + N_G + 5$ Majorana fermion, while the
ghost and ghostino contributions are equivalent to, respectively,
$-2$ times the contribution of a $(4+N_G)$-component scalar and
$+2$ times the contribution of a four-component scalar, such that
supersymmetry of the off-shell spectrum is maintained. The
matrix-valued covariant derivatives $D_{\Phi}$ and $D_{\Theta}$,
metric factors $Z_{\Phi}$ and $Z_{\Theta}$, as well as the
quantities $H_{\Phi}$ and $M_{\Theta}$ are defined
in~\cite{Gaillard:1993es}. With the gauge-fixing prescription
described in that work, the one-loop contribution to the effective
action is
\begin{eqnarray}
\Lag_1 &=& \frac{i}{2}\Tr\ln(\hat{D}_{\Phi}^{2} + H_{\Phi}) -
\frac{i}{2}\Tr\ln(-i\notD_{\Theta} + M_{\Theta}) + i\Tr\ln(D_{\rm
gh}^{2} + H_{\rm gh}) -i\Tr\ln(D_{\rm Gh}^{2} + H_{\rm Gh})
\nonumber \\ &\equiv& \frac{i}{2}\STr\ln(\hat{D}^{2} + H) + T_{-}
.
\label{Lag1} \end{eqnarray}
To obtain the second line of~(\ref{Lag1}) we have split the
fermionic contribution into a helicity-even piece given by
\begin{equation}
\hat{D}_{\Theta}^{2} + H_{\Theta} \equiv (-i\notD_{\Theta} +
M_{\Theta})(i\notD_{\Theta} + M_{\Theta})
\end{equation}
and a helicity-odd contribution $T_{-}$ which contains no
quadratic divergences. We will therefore neglect this contribution
in what follows.

If we were to explicitly evaluate the quantities in~(\ref{Lag1}),
using an ultraviolet cut-off $\Lambda$ for the momentum
integration in
\begin{equation}
\int d^4 x \;  d^4 p \; \STr\ln(p^2 + H) ,
\label{RAWint} \end{equation}
we would obtain the quadratic divergence
\begin{equation}
\Lag_1^{\QUAD} = \sqrt{g}\frac{\Lambda^2}{32\pi^2}\STr H ,
\label{Lquad} \end{equation}
where the supertrace includes a trace over bosonic, fermionic,
ghost and ghostino degrees of freedom. As we will be ultimately
interested in effective Lagrangians describing superstring
theories, we allow for noncanonical gauge field kinetic energy by
coupling the Yang-Mills sector of the theory to a holomorphic
function of chiral multiplets via the gauge kinetic function
$f_{ab} = \delta_{ab} f$ with $f=x+iy$. Then the quadratically
divergent contributions from the (gauge fixed) gravity sector, the
$N_{\chi}$ chiral multiplets and the (gauge fixed) Yang-Mills
sector of internal symmetry dimension $N_G$ are
%
\begin{eqnarray}
\STr H_{\rm grav} &=& -10V -2e^{-K}A\oline{A} + \frac{7}{2}r +
4K_{i\bar{m}} {\cal D}_{\mu} z^i {\cal D}^{\mu}\bar{z}^{\bar{m}} -
\frac{x}{2}F_{\mu\nu}^a F^{\mu\nu}_a -\frac{f_i \oline{f}^i}{2x^2}
\mathcal{D} , \nonumber \\
\STr H_{\chi} &=& 2N_{\chi} \(\wh{V} + e^{-K}A\oline{A} -
\frac{1}{4}r\) +\frac{f_i \oline{f}^i}{2x^2} \mathcal{D} +
\frac{2}{x}{\cal D}_a D_i(T^az)^i
-2R_{i\bar{m}}\(e^{-K}\oline{A}^iA^{\bar{m}} + {\cal D}_{\mu} z^i
{\cal D}^{\mu}\bar{z}^{\bar{m}}\) ,
\nonumber \\
\STr H_{\rm YM} &=& \frac{N_G}{2}r + \frac{1}{x}{\cal D}_a{\cal
D}^a + \frac{x}{2}F_{\mu\nu}^a F^{\mu\nu}_a
-\frac{N_G}{2x^2}e^{-K} f_i \bar{f}^j A_j \bar{A}^i
-\frac{N_G}{2x^2}\(\partial_{\mu}x\partial^{\mu}x +
\partial_{\mu}y \partial^{\mu}y\).
\label{Hparts} \end{eqnarray}
The tree-level scalar potential is given by $V = \wh{V} +
\mathcal{D}$ with
\begin{equation}
\wh{V} = e^{-K}(A_i \oline{A}^i -3A\oline{A}) \; ; \quad
\mathcal{D} = \frac{1}{2x}\mathcal{D}_a\mathcal{D}^a \; , \quad
\mathcal{D}_a = K_i(T^a z)^i.
\label{potential} \end{equation}

Combining all of these we get the total light field contribution
\begin{eqnarray}
\STr H &=& -\(N_{\chi} -N_G -7\)\frac{r}{2} + 2(N_{\chi} -5)V
-(N_{\chi}-1)\frac{1}{x}{\cal D}_a{\cal D}^a + \frac{2}{x}{\cal
D}_a D_i(T^az)^i \nonumber \\
 & & + 2(N_{\chi} -1)e^{-K}A\oline{A} -2(R_{i\bar{m}} -2K_{i\bar{m}}) {\cal
D}_{\mu} z^i {\cal D}^{\mu}\bar{z}^{\bar{m}}
-2R_{i\bar{m}}e^{-K}\oline{A}^iA^{\bar{m}} \nonumber \\
 & & -\frac{N_G}{2x^2}e^{-K} f_i \bar{f}^j A_j \bar{A}^i -\frac{N_G}{2x^2}
 \(\partial_{\mu}x\partial^{\mu}x + \partial_{\mu}y
 \partial^{\mu}y\) ,
\label{Hlight} \end{eqnarray}
which we can write in a slightly more compact and suggestive
manner\footnote{In arriving at~(\ref{Hlight}) and~(\ref{Hlight2})
we have corrected some errors in expressions found
in~\cite{Gaillard:1996hs,Gaillard:1994sf}. These corrections are
summarized in Appendix~B and can also be found
in~\cite{Gaillard:1998bf,Gaillard:1999ir}.} by writing $f_i
\partial_{\mu} z^i = i\partial_{\mu} x + i\partial_{\mu} y$ and by
noting that $F^i = -e^{-K/2}\oline{A}^i =
-e^{-K/2}K^{i\bar{m}}\oline{A}_{\bar{m}}$:
\begin{eqnarray}
\STr H &=& -\(N_{\chi} -N_G - 7\)\frac{r}{2} - 8\mathcal{D} + (28
-4N_{\chi}) e^{-K}A\oline{A} +2K_{i\bar{m}}\[2{\cal D}_{\mu} z^i
{\cal D}^{\mu} \bar{z}^{\bar{m}} + (N_{\chi}-5)F^i
\oline{F}^{\bar{m}}\] \nonumber \\
& & -2\(\frac{N_G}{4x^2} f_i \bar{f}_{\bar{m}} +
R_{i\bar{m}}\)({\cal D}_{\mu} z^i {\cal D}^{\mu} \bar{z}^{\bar{m}}
+ F^i \oline{F}^{\bar{m}}) + \frac{2}{x}{\cal D}_a D_i(T^az)^i .
\label{Hlight2} \end{eqnarray}
Note that by making the following identifications
\begin{equation}
\lang e^{-K}A\oline{A} \rang = m_{3/2}^{2} \; ; \quad
\lang \frac{f_i \bar{f}_{\bar{m}}}{4x^2} F^i \oline{F}^{\bar{m}}
\rang = m_{1/2}^{2} \; ; \quad
\lang K_j^k e^{-K}A\oline{A} - R_{ji\bar{m}}^k F^i
\oline{F}^{\bar{m}} \rang = (m^2)_j^k
\label{soft} \end{equation}
we recognize a contribution to the vacuum energy analogous to that
given in~(\ref{STr}).

A subset of these terms were considered in~\cite{Choi:1997de}
where the potential impact on soft terms was investigated. Yet the
one-loop Lagrangian~(\ref{Lquad}), with traces given
by~(\ref{Hparts}), is {\em not yet in a suitable form for
extracting information on the resulting low-energy theory}. Most
obviously, the loop contribution to the Einstein term in the
gravity action
\begin{equation}
\Lag = \[ 1 - \frac{\Lambda^2}{32\pi^2}(N_{\chi} - N_G
-7)\]\frac{r}{2} + \dots
\label{rquad} \end{equation}
indicates that the theory cannot be consistently expanded about a
flat metric until a Weyl rescaling is performed to render this term
canonical. But this straightforward concern is not the only one. We
expect our supergravity theory to represent merely an effective
Lagrangian, completed at some high energy scale by an underlying --
and presumably finite -- theory, such as string theory. Thus, the
apparent divergences must be rendered finite by the inclusion of an
appropriate regulating sector (with typical mass scale $\mu$ as
in~(\ref{subtract})) before any low-energy phenomenology can be
computed~\cite{Binetruy:1988dw}. Low-energy results will depend on
the $\eta_S$ and this can only be obtained by employing a
regularization scheme consistent with local supersymmetry (as well
as the known symmetries of the underlying theory). That is, the
coefficients of the quadratically divergent terms are unreliable in
the absence of a manifestly supersymmetric regularization procedure.

\subsection{Rendering the Divergence Finite \`a la Pauli-Villars}
\label{sec:PV} We now return to~(\ref{Lag1}) but this time we include in the
traces the contributions from a Pauli-Villars regulating
sector~\cite{Gaillard:1994sf,Gaillard:1998bf,Gaillard:1999ir}. We
separate the effective PV mass into two contributions
\begin{equation}
M_{\PV}^{2} = H_{\PV}(\phi) + \( \begin{array}{cc} \mu^2 & \nu \\
\nu^{\dagger} & \mu^2 \end{array} \) \equiv H_{\PV} + \mu^2 + \nu
.
\label{HPV} \end{equation}
The first contribution, $H_{\PV}(\phi)$, is the analog to the
objects labeled $H$ for the light fields of the theory. It is a
field-dependent quantity for which any mass scale is only
implicit, via the $vev$ of one or more light fields. But in
addition we have true supersymmetric mass terms ({\em i.e.} terms
that arise from superpotential couplings) that we introduce as
well.\footnote{While these terms have explicit mass scales
associated with them, they may still retain some residual
field-dependence. For example, maintenance of modular invariance
in the regulated low-energy theory will often require that these
mass terms have some K\"ahler modulus dependence.} We assume these
masses to represent the scale of the UV-completion of the
supergravity theory, and thus $|\nu|^2 \sim \mu^2 \gg H_{\PV} \sim
H$.

Combining the field-dependent contributions of the light and heavy
(PV) fields, we define the matrix $H' = H + H^{\PV}$ and now
re-express the integral~(\ref{RAWint}) with these contributions
included
\begin{equation}
S_1^{\QUAD} = \frac{1}{32\pi^2} \int \diff^4x \; \diff^4p \;
\sqrt{g} \; \STr \ln \(p^2 + H' + \mu^2 + \nu\) .
\label{S1integral} \end{equation}
Expanding the logarithms in~(\ref{S1integral}) in powers of
$(H'+\nu)/(p^2+\mu^2)$, and using the fact that $H' \ll \nu \ll
\mu^2$ with $\Tr \; \nu =0$, the momentum integration can be
performed. Requiring that the coefficient of the $\Lambda^2$ term
(the quadratic divergence) vanishes implies that
\begin{equation}
\STr(\mu^2) = \STr H' = 0
\label{quadcond} \end{equation}
while requiring the coefficient of the $\ln \Lambda^2$ term (the
logarithmic divergence) vanishes implies additional
conditions~\cite{Gaillard:1994sf,Gaillard:1998bf,Gaillard:1999ir}.
The vanishing of $\STr (\mu^{2n})$ is automatically ensured by
supersymmetry. Once all conditions are satisfied, the momentum
integration is rendered finite and the resulting one loop
contribution is now proportional to the square of the explicit PV
masses in~(\ref{HPV})
\begin{equation}
S_1^{\QUAD} = - \int \frac{\diff^4x}{64\pi^2}\sqrt{g} \;
\STr\[\(2\mu^2 H' + \nu^2\)\ln\mu^2\] \ + {\cal O}(\ln\mu^2) .
\label{S1final} \end{equation}

To obtain the explicit forms of the matrices $H_{\PV}$, $\mu$ and
$\nu$ we must specify the Pauli-Villars field content. The
regulation of matter and Yang-Mills loop contributions to the
matter wave function renormalization requires the introduction of
PV chiral superfields $\Phi^P =
\Phi^I_{\alpha},\wh{\Phi}^I_{\alpha},\Phi^a$, which transform
according to the chiral matter, anti-chiral matter and adjoint
representations of the gauge group and have signatures $\eta^P_p =
-1,+1,+1,$ respectively for modes $p$ labeled collectively by $P$.
Note that full regulation of the theory requires $\alpha$ copies
of chiral fields with the same gauge quantum numbers as the light
fields $Z^i$. These fields couple to the light fields through the
superpotential
\begin{equation}
W(\Phi^P,Z^i) =
\frac{1}{2}\sum_{\alpha}W_{ij}(Z^k)\Phi^I_{\alpha}\Phi^J_{\alpha}
+ \sqrt{2}\sum_{\alpha}\Phi^a \wh{\Phi}^I_{\alpha}(T_aZ)_i +
\cdots \label{PVcoup}
\end{equation}
where $T_a$ is a generator of the gauge group $\mathcal{G}_a$. The
K\"ahler potential for these fields can be written in the general
form
\begin{equation}
K(\Phi^P, \oline{\Phi}^{P} ) =
\kappa^{\Phi}_{I\bar{M}}\Phi^I_{\alpha}
\oline{\Phi}^{\bar{M}}_{\alpha} + \wh{\kappa}^{\Phi}_{I\bar{M}}
\wh{\Phi}^I_{\alpha} \wh{\oline{\Phi}}^{\bar{M}}_{\alpha} +
\kappa^{\Phi}_a |\Phi^a|^2
\label{pvkal} \end{equation}
where the functions $\kappa^{P}$ are {\em a priori} functions of
the hidden sector (moduli) fields. The PV mass for each superfield
$\Phi^P$ is generated by coupling it to another field
$\Pi^{P}=(\Pi^I, \widehat{\Pi}^I, \Pi^a)$ in the representation of
the gauge group conjugate to that of $\Phi^P$ through a
superpotential term \begin{equation} W_m =
\sum_{p}\mu_{PQ}(Z^n)\Phi^P \Pi^Q, \label{PVbilinear}
\end{equation} where $\mu_{PQ}(Z^n)$ can in general be a
holomorphic function of the light superfields. This is the origin
of the explicit $\mu$ and $\nu$-dependent mass terms
in~(\ref{HPV}).

These regulator fields $\Phi^P$ must be introduced in such a way
as to cancel the quadratic divergences of the light field loops --
and thus their K\"ahler potential is determined relative to that
of the fields which they regulate. Specifically we have
\begin{equation}
\lbr \begin{array}{l} \Phi^I:\;\;\kappa^{\Phi}_{I\bar{M}} =
\kappa_{i\bar{m}} = K_{i\bar{m}} \\
\wh{\Phi}^I:\;\; \h{\kappa}^{\Phi}_{I\bar{M}} =
\kappa^{-1}_{i\bar{m}} = K^{i\bar{m}} \\
\Phi^a: \;\;\kappa^{\Phi}_a \; \; = g_a^{-2}e^K
\end{array} , \right.
\label{kapPhi}
\end{equation}
where $g_a$ is the (possibly field-dependent) gauge coupling
constant for the gauge subgroup $\mathcal{G}_a$. There is no
similar constraint on the K\"ahler potential for the fields
$\Pi^P$, but this uncertainty plays no significant role in an
examination of the contributions to the scalar potential in which
we are interested.\footnote{See, for
example,~\cite{Gaillard:2000fk,Binetruy:2000md} for the impact of
these fields on logarithmic contributions to the one-loop scalar
potential.}  For concreteness, in the following we set
\begin{equation}
\lbr \begin{array}{l} \Pi^I:\;\;\kappa^{\Pi}_{I\bar{M}} =
\delta_{i\bar{m}} e^{\alpha^I K} \\
\wh{\Pi}^I:\;\; \h{\kappa}^{\Pi}_{I\bar{M}} = K_{i\bar{m}}
e^{\hat\alpha^I K}
\\
\Pi^a: \;\;\kappa^{\Pi}_a \; \; = g_a^2 e^{(\alpha^a -1)K} .
\end{array} \right.
\label{kapPi}
\end{equation}

To regulate the quadratically divergent terms arising from the
non-canonical nature of the gauge kinetic energy -- such as the
last term in~(\ref{Hlight}) -- we will here take a very simple PV
sector in which we add $N_G$ chiral multiplets $\varphi^{\alpha}$
for each gauge group factor, with a universal K\"ahler potential
coupling
\begin{equation}
K(\varphi, \oline{\varphi}) = \kappa^{\varphi}_{\alpha}
|\varphi^{\alpha}|^2\; ; \quad \kappa^{\varphi}_{\alpha} = (f +
\oline{f})  = 2g_{\STR}^{-2},
\end{equation}
where $f = x+iy$ is the field-dependent coefficient of the gauge
kinetic function introduced earlier, and $g_a = g_{\STR}$ at the
string scale. The supersymmetric mass term for this set of fields
arises from the following term in the superpotential
\begin{equation} W(\varphi^\alpha) =
\frac{1}{2}\sum_\alpha \mu_{\varphi}^{\alpha} (\varphi^\alpha)^2 .
\end{equation}
If we were to consider logarithmic divergences, including
dilaton-dependent terms that arise from gauge loops, we would
introduce a different set of chiral superfields with different
K\"ahler potential couplings~\cite{Gaillard:1999ir}, but this is
sufficient for our purposes.

In addition to these chiral superfields, the regulation of the
term in~(\ref{Hlight2}) proportional to the curvature $r$ requires
the introduction of $U(1)_b$ gauge multiplets $W^b$ with signature
$\eta^b$. As these regulator fields must have a supersymmetric
mass, we must introduce a chiral superfield $\Phi^b =
e^{\theta_b}$, with the same signature, which will eventually form
a massive vector supermultiplet with the $W^b$ fields. The
K\"ahler potential for these chiral fields is given by
\begin{equation}
K(\theta,\oline{\theta}) = \frac{1}{2} \sum_b \nu_b e^{\alpha_b K}
(\theta_b + \oline{\theta}_b)^2 .
\label{Ktheta} \end{equation}
This K\"ahler potential has an invariance under this $U(1)$ for
which the field $\Phi^b$ has charge $q_b \delta_{bc}$ and where
$\delta_c \theta_b = i q_b \delta_{bc}$. The corresponding D-term
\begin{equation}
\mathcal{D}(\theta, \oline{\theta}) = \frac{1}{x_b} \mathcal{D}_b
\mathcal{D}^b \; ; \quad \mathcal{D}_b = \sum_c K_c
\delta_b\theta^c = i (\theta^b + \oline{\theta}^b)q_b e^{\alpha_b
K} \nu_b
\label{Dtheta} \end{equation}
vanishes in the background (where we set all Pauli-Villars fields
to zero), but the combination $(\theta^b +
\oline{\theta}^b)/\sqrt{2}$ acquires a supersymmetric squared mass
\begin{equation}
\mu_b^2 = \frac{1}{2x_b} q_b^2 e^{\alpha_b^{\theta} K} \nu_b
\label{thetamass} \end{equation}
which is equal to the mass of the vector superfields $W_b$, with
which it forms a massive vector multiplet in accordance with the
Higgs effect.\footnote{$x_b = \re f_b(z)$, where $f_b(Z)$ is the
gauge kinetic function for $W^b$. Regulation of logarithmic
divergences from both dilaton and gravity loops requires fields
with both $f_b(Z) =$ constant$\times S$ and $f_b(Z) =$ constant.
This in unimportant here and for simplicity we take $f_b(Z) = $
constant.}

Having collected all the elements we need for regulating the
quadratic divergences we can compute the Pauli-Villars
contribution to the supertraces, and thus $H' = H + H^{\PV}$:
\begin{eqnarray}
\STr H' &=&  2\wh{V}\[N_{\chi}\(1+\sum_\alpha\eta^I_\alpha\) +
\sum_P\eta_P\(1 - \alpha_P\) + \sum_b\eta_b\(1 - \alpha_b\) - 5\]
\nonumber \\
& & + 2e^{-K}A\oline{A}\[N_{\chi}\(1+\sum_\alpha\eta^I_\alpha\) +
\sum_P\eta_P\(1 - 3\alpha_P\) + \sum_b\eta_b\(1 - 3\alpha_b\) -
1\] \nonumber \\
& & - \frac{r}{2}\[N_{\chi}\(1+\sum_\alpha\eta^I_\alpha\) +
\sum_P\eta_P - 7 - N_G\] - 2
R_{i\bar{m}}\(e^{-K}\oline{A}^iA^{\bar{m}} + \D_\mu
z^i\D^\mu\bar{z}^{\m}\)\(1+\sum_\alpha\eta^I_\alpha\) \nonumber
\\
& & + 2\(K_{i\bar{m}}\D_\mu z^i\D^\mu\bar{z}^{\bar{m}} - 2\D\)\(2
- \sum_P\eta_P\alpha_P - \sum_b\eta_b\alpha_b\) \nonumber \\
& & + \(\sum_{\alpha} \eta_{\alpha}^{\varphi} -N_G\)
\[\frac{1}{2x^2} f_i \bar{f}_{\bar{m}} \oline{F}^i F^{\bar{m}}
+\frac{1}{2x^2}\(\partial_{\mu}x\partial^{\mu}x + \partial_{\mu}y
 \partial^{\mu}y\)\],
\label{Hprime} \end{eqnarray}
where the summation over the index $P$ is a shorthand for summing
over all of the fields $\Phi^a$, $\wh{\Phi}^I$, $\Pi^A$ and
$\varphi^{\alpha}$. Note that $\alpha_P =0$ when $P$ represents
the field $\varphi^{\alpha}$. Now given that~(\ref{Hprime}) is the
coefficient of the quadratic divergence in~(\ref{S1final}) it is
natural to insist that each term in the expression be separately
vanishing, as required by~(\ref{quadcond}). That would imply the
following relations among signatures and K\"ahler factors
$\alpha$:
\begin{equation}
0 = 1+\sum_\alpha\eta^I_\alpha = \sum_P\eta_P + \sum_b\eta_b - 7 =
\sum_P\eta_P - 7 - N_G = 2 - \sum_P\eta_P\alpha_P -
\sum_b\eta_b\alpha_b = N_G - \sum_{\alpha}\eta_{\alpha}^{\varphi}.
\end{equation}
This is certainly reasonable in the case where one considers only
the quadratic divergences of the theory, as
in~\cite{Gaillard:1994sf}, since an off-shell regularization is
possible in this case. That is to say, the theory can be
regularized even in the case where the Einstein term is not
canonically normalized, as is clearly the case for the curvature
term in~(\ref{Hprime}). However, in a complete treatment that
includes the logarithmic divergences it is necessary to do a Weyl
rescaling at this stage prior to imposing the constraint
of~(\ref{quadcond}). More specifically, a Weyl transformation
removes a term proportional to the linear combination $\half r +
K_{i\bar{m}}\D_{\mu} z^i\D^{\mu}\bar{z}^{\bar{m}} - 2V$, which
vanishes on shell due to the graviton equations of motion. Thus we
are only requiring on-shell finiteness. Of course, in this case
the vanishing of the logarithmic divergences will impose
additional constraints on the signatures $\eta$ and K\"ahler
factors $\alpha$.


Having imposed these constraints, we are now in a position to
evaluate the terms quadratic in the Pauli-Villars masses
in~(\ref{S1final}). We can simplify the expressions by factoring
out of the mass terms any dependence on light fields via $\mu_p =
\beta_p \mu_P(z)$ and $\nu_b = x_b
(\beta_b^{\theta})^2|\mu_{\theta}(z)|^2$. Without loss of
generality we will set $q_b =1$ in~(\ref{thetamass}) and take
$\alpha_a \equiv \alpha_{\Phi}$, $\alpha_b \equiv \alpha_{\theta}$
and $\beta_b \equiv \beta_{\theta}$ to be independent of $a$ and
$b$, respectively. Then the relevant terms are
\begin{eqnarray}
\STr(2\mu^2 H' + \nu^2) &=&
e^{-K}A_{IJ}\oline{A}^{IJ}\[K_{i\bar{m}}\D_\mu\bar{z}^{\bar{m}}\D^\mu
z^i + 3e^{-K} A\oline{A}\] +
2e^{-2K}A_{IJ}\oline{A}^{JK}R^{m\;\;I}_{\;\;n\;\;K}A_m\oline{A}^n
\nonumber \\
& & + e^{-2K}\[A_{kIJ}\oline{A}^{IJm}\oline{A}^k A_m - (A_{IjK}
\oline{A}^{IK}\oline{A}^j A + \hc )\] - \frac{e^{-K}}{x}
\D_a(T^az)^i A_{iJK}\oline{A}^{JK} \nonumber \\
& & +\D_\mu\bar{z}^{\bar{m}}\D^\mu z^i
e^{-K}\(A_{iJK}\oline{A}^{JK}_{\bar{m}}
 + 2R^K_{i\bar{m} J}e^{-K}A_{KL}\oline{A}^{JL}\) \nonumber \\
& & +4\sum_b  e^{\alpha^{\theta}_b K} |\beta_b \mu_{\theta}|^2 \[
\(\wh{V} +
 e^{-K}A\oline{A}\) -{e^{-(1+\alpha^{\theta}_b)K}}|A-(\alpha^{\theta}_b
 K_{\bar{m}} + \partial_{\bar{m}} \ln\oline{\mu}_{\theta})
 A^{\bar{m}}|^2 \right. \nonumber \\
& & \left. - \D_\mu\bar{z}^{\bar{m}}\D^\mu z^i (\alpha^{\theta}_b
K_{\bar{m}} + \partial_{\bar{m}}
\ln\oline{\mu}_{\theta})(\alpha^{\theta}_b K_{i} + \partial_{i}
\ln\oline{\mu}_{\theta} )\right. \nonumber \\
& & \left.
-\alpha^{\theta}_bK_{i\bar{m}}\(\D_\mu\bar{z}^{\bar{m}}\D^\mu z^i
+ A^{\bar{m}}\oline{A}^i\) + 2 \alpha_b^{\theta} \mathcal{D}\] ,
\label{everything} \end{eqnarray}
where we have already inserted the explicit matrix elements for
the $\theta_b$ fields. The upper-case indices then represent each
of the fields we had labeled with $P$ in~(\ref{Hprime}): the
$\Phi^A$ as well as the $\varphi$ fields. Lower-case indices refer
to light fields and indices of both types are raised with the
inverse metric for the appropriate field. For completeness we
collect the relevant matrix elements for all the fields below:
\begin{eqnarray}
\Phi^I,\Pi^I &:&
\sum_{I,J}e^{-K}A_{(I\alpha)(J\alpha)}\oline{A}^{(K\alpha)(J\alpha)}
= 2\delta^K_I\sum_{i=I}e^{K(1 - \alpha_\alpha)}
K^{i\ibar}|\beta_{\alpha}\mu|^2 \nonumber
\\
 & & A_{(I\alpha)(J\alpha)k} = \[\(1 - \alpha_\alpha\)K_k -
\partial_k\ln\mu \]A_{(I\alpha)(J\alpha)} - \Gamma^{\ell}_{ki}
A_{(L\alpha)(J\alpha)} \nonumber \\
\wh{\Phi}^{I},\wh{\Pi}^{I} &:&
\sum_Je^{-K}A_{(I\alpha)(J\alpha)}\oline{A}^{(K\alpha)(J\alpha)} =
2\delta^K_I e^{-\wh{\alpha}_{\alpha}
K}|\wh{\beta}_{\alpha}\wh{\mu}|^2 \nonumber \\
 & & A_{(I\alpha)(J\alpha)k} =  - \(\wh{\alpha}_{\alpha} K_k +
\partial_k\ln\wh{\mu} \)A_{(I\alpha)(J\alpha)} \nonumber \\
\Phi^a,\wh{\Phi^a} &:& e^{-K}A_{cb}\oline{A}^{ac} = 2\delta^a_b
e^{-\alpha^{\Phi}_a K} |\beta_a^{\Phi}\mu_\Phi|^2 \nonumber \\
 & & A_{ab\;i} = -\(\alpha^{\Phi}_a K_i +
 \partial_i\ln\mu_{\Phi}\)A_{ab}\nonumber \\
\varphi^{\alpha} &:& e^{-K}A_{\gamma\beta}\oline{A}^{\alpha\gamma}
= \frac{e^K}{4x^2}\delta^{\alpha}_{\beta}
|\beta_{\alpha}^{\varphi}\mu_{\varphi}|^2 \nonumber
\\
 & & A_{\alpha\beta i} = \[K_i - \frac{f_i}{x} -
 \partial_i\ln\mu_{\varphi}\]A_{\alpha\beta} \nonumber \\
\theta^b &:& \frac{1}{x_b}\sum_{c,d} \delta^b \theta^d
K_{d\bar{c}} \delta_b \oline{\theta}^{\bar{c}} =
e^{\alpha^{\theta}_b K} |\beta_{b}^{\theta}\mu_{\theta}|^2
\nonumber \\
 & & A_{bc} = \nu_b e^{\alpha^{\theta}_b K} A_{c}^{\bar{b}}
 = \nu_b e^{\alpha^{\theta}_b K} \[A - (\alpha^{\theta}_b K_{\bar{m}}
 + \partial_{\bar{m}}\ln \oline{\mu}_{\theta})A^{\bar{m}}\] \delta_{bc} .
\label{matrix} \end{eqnarray}
The relevant terms for the scalar reparameterization connection
$\Gamma$ and the associated Riemann tensor $R$ are given by
\begin{eqnarray}
(\Gamma_\Phi)^a_{bk} &=& - (\Gamma_\Pi)^a_{bk} +
\alpha^{\Phi}_{a}\delta^a_b K_k = \delta^a_b\(K_k - \partial_k\ln
g_a\) \nonumber \\
(R_\Phi)^a_{b k\bar{m}} &=& - (R_\Pi)^a_{b k\bar{m}} +
\alpha^{\Phi}_a \delta^a_b K_{k\bar{m}} = \delta^a_b\(K_k -
\partial_{\bar m}\pp_k\ln g_a\), \nonumber \\
(\Gamma_\Phi)^{(I\alpha)}_{(J\beta),k} &=& -
(\Gamma_{\wh{\Phi}})^{(J\alpha)}_{(I\beta),k} =
(\Gamma_{\wh{\Pi}})^{(I\alpha)}_{(J\beta),k} -
\delta^\alpha_\beta\delta^I_J \hat\alpha_{\beta} K_k =
\delta^\alpha_\beta\Gamma^i_{jk}, \nonumber\\
(R_\Phi)^{(I\alpha)}_{(J\beta),k\bar{m}} &=& -
(R_{\wh{\Phi}})^{(J\alpha)}_{(I\beta),k\bar{m}} =
(R_{\wh{\Pi}})^{(I\alpha)}_{(J\beta),k\bar{m}} -
\delta^\alpha_\beta\delta^I_J \hat\alpha_{\beta} K_{k\m}
=\delta^\alpha_\beta R^i_{jk\bar{m}}, \nonumber \\
(\Gamma_\Pi)^{(I\alpha)}_{(J\beta),k} &=&
\delta^\alpha_\beta\delta^I_J\alpha_{\alpha} K_k, \;\;\;\;
(R_\Pi)^{(I\alpha)}_{(J\beta),k\bar{m}} = \delta^\alpha_\beta
\delta^I_J \alpha_{\alpha} K_{k\m},
\nonumber \\
(\Gamma_\varphi)^{\beta}_{\alpha i} &=&
\delta_{\alpha}^{\beta}\frac{f_i}{2x}, \;\;\;\;
(R_\varphi)^{\beta}_{\alpha i\bar{m}} = -\delta_\alpha^\beta
\frac{f_i \bar{f}_{\bar{m}}}{4x^2},
\nonumber \\
(\Gamma_\theta)^{c}_{ib} &=& \delta_{b}^{c}(\alpha^{\theta}_b K_i
+ \partial_i \ln \mu_{\theta}), \nonumber\\
(R_\theta)_{b\bar d i\bar{m}} &=& -\delta_{b d}e^{\alpha_b^\theta
K}\nu_b\alpha_b^{\theta}e^{\alpha_b{\theta} K} K_{i\bar{m}} =
-\delta _{d c}e^{\alpha_b^\theta K}\nu_b R^{c}_{b i\bar{m}}
\label{connections} \end{eqnarray}
%

\subsection{Renormalization of the K\"ahler Potential}
\label{sec:renorm} These, then, are the terms quadratically dependent on the
(presumably large) cut-off scale $\mu_{\PV}$. We now have the
necessary ingredients to study the phenomenological implications
of these terms. Before turning our attention there, however, it is
instructive to consider how these terms can be grouped into a
renormalization of the space-time metric and the K\"ahler
potential itself. Let us define the (field-dependent) effective
cut-offs in terms of the Pauli-Villars mass terms in the following
way
\begin{eqnarray}
\Phi^I,\Pi^I &:& e^{(1
-\alpha_\alpha)K}K^{i\ibar}|\beta_{\alpha}\mu|^2 \equiv
\(\beta_{\alpha}\)^2 \Lambda_\alpha^2 \nonumber \\
\wh{\Phi}^{I},\wh{\Pi}^{I} &:& 2e^{-\hat\alpha_\alpha K}
|\hat\beta_\alpha\hat\mu|^2 \equiv \(\hat\beta_{\alpha}\)^2
\wh\Lambda_\alpha^2 \nonumber \\
\Phi^a,\Pi^a &:& 2e^{-\alpha_\alpha^\Phi K}
|\beta_\alpha^\Phi\mu_\Phi|^2 \equiv
(\beta_\alpha^\Phi)^2\Lambda_{\Phi}^2 \nonumber \\
\varphi^{\alpha} &:& \frac{e^K}{4x^2}
|\beta^\varphi_{\alpha}\mu_\varphi|^2 \equiv
(\beta^\varphi_{\alpha})^2 \Lambda_{\varphi}^2 \nonumber \\
\theta^b &:&  2e^{\alpha_b^{\theta} K}
|\beta_{b}^\theta\mu_{\theta}|^2 \equiv (\beta_b^\theta)^2
\(\Lambda_b^{\theta}\)^2 .
\label{lambdas} \end{eqnarray}
With this compact notation the quadratic dependence on the PV mass
can be easily grouped into a correction of the K\"ahler potential
and space-time metric. In other words, defining $S_0 + S_1 = \int
\diff^4 x \;(\Lag_0 + \Lag_1)$ we have
\begin{equation}
\Lag_0(g_{\mu\nu}^0, K) + \Lag_1 = \Lag_0(g_{\mu\nu}, K+\delta
K)\; , \quad g_{\mu\nu} = g_{\mu\nu}^0(1+\epsilon) .
\label{Lagrenorm} \end{equation}
The above expression involves a summation over effective cutoffs
$\Lambda_P$ that runs over {\em all} the heavy PV modes $p$
labeled by $P$
\begin{equation}
\epsilon = -\sum_P \lambda_P \; \zeta'_P
\frac{\Lambda_P^2}{32\pi^2} \; ; \quad \delta K = \sum_P \lambda_P
\; \zeta_P \frac{\Lambda_P^2}{32\pi^2}
\label{deltas} \end{equation}
where
\begin{equation}
\lambda_P = \sum_p \eta_p^P (\beta_p^P)^2 \ln (\beta_p^P)^2
\label{lambdaP} \end{equation}
sums over the various species within each class, and the
coefficients $\zeta$ and $\zeta'$ are given by
\begin{eqnarray}
\Phi^I; \; \wh{\Phi}^{I} &:& \zeta_I = 1 \; ; \quad \zeta'_I = 1
\nonumber \\
\Phi^a &:& \zeta_a = 1 \; ; \quad \zeta'_a = 1 \nonumber \\
\varphi^{\alpha} &:& \zeta_{\varphi} =1 \; ; \quad
\zeta'_{\varphi} = 1 \nonumber
\\
\theta^b &:&  \zeta_b = -4 \; ; \quad \zeta'_b = 0.
\label{zetas} \end{eqnarray}

These are precise expressions for the quantities alluded to in
equations~(\ref{subtract}) in Section~\ref{sec:general} above. Let
us see how the identification in~(\ref{Lagrenorm})
and~(\ref{deltas}) comes about through an explicit construction.
Define the quantity $\wtd{K} = K + \delta K$ where
\begin{equation}
\delta K = \sum_P\delta K_P, \qquad \delta K_P = \zeta_P\lambda_P
\frac{\Lambda^2_P}{32\pi^2} =
\loopp\sum_p\zeta_P\eta_p\ln(\beta^2_p)\delta k_p
\label{tildek} \end{equation}
and the field dependence of the cut-offs in~(\ref{lambdas}) is
here represented by
\begin{equation}
\delta k_p = \beta_p^2 \Lambda_P^2 .
\end{equation}
The content of~(\ref{Lagrenorm}) can be expressed as an expansion
in $\delta K$ as follows~\cite{Gaillard:1996ms}
\begin{eqnarray}
\frac{1}{\sqrt{g}}\Lag_0\(\wtd{K}\) &=&
\frac{1}{\sqrt{g}}\Lag_0(K) - \delta K\wh{V} + \delta
K_{i\jbar}\(\D^{\mu} z^i\D_{\mu}\bar{z}^{\jbar} +
F^i\bar{F}^{\jbar}\) \nonumber \\
 & &-\(\delta K_i\[F^i\mG +
\frac{1}{2x}\D_a(Tz)^i\]+\hc\) \nonumber
\\
 &\equiv& \frac{1}{\sqrt{g}}\Lag_0(K) + \loopp\sum_P\zeta_P
\sum_p\eta_p\ln(\beta^2_p)\ell_p.
\label{Lexpand} \end{eqnarray}
To determine the values of the various $\ell_p$ it is necessary to
take covariant derivatives of the $\delta k_p$. Using the fact
that the metric is covariantly constant: $D_i K^{IJ} = 0$, and the
relations
\begin{eqnarray} A_{I J k} &=& D_k A_{I J} = \partial_k A_{I J} -
\Gamma^L_{I k}A_{L J} - \Gamma^L_{J k}A_{I L}, \qquad
D_k(e^{-K}\oline{A}_{\bar I\bar J}) = 0, \nonumber \\
D_{\bar{m}}(e^{-K}D_k A_{I J}) &=& e^{-K}(K_{k\bar{m}}A_{I J} -
R^L_{I k\bar{m}}A_{L J} - R^L_{J k\bar{m}}A_{I L}),
\label{daij}\end{eqnarray}
we have for $P \neq \theta$
\begin{eqnarray}
\delta k_p &=& \beta_p^2\Lambda^2_P = e^{-K}A_{I J}\oline{A}^{I J}
\nonumber \\
 D_k\delta k^p &=& \delta k^p_k = D_k(e^{-K}A_{I J}\oline{A}^{I
J})_p = e^{-K}(A_{I J k}\oline{A}^{I J})_p, \nonumber \\
D_{\bar{m}}D_k\delta k^p &=& \delta k^p_{k\bar{m}} = e^{-K}(A_{I J
k}\oline{A}^{I J}_{\bar{m}} + K_{k\bar{m}} A_{I J}\oline{A}^{I J}
- 2R^L_{I k\bar{m}}A_{L J}\oline{A}^{I J})_p.
\label{ddelk}\end{eqnarray}
From~(\ref{Lexpand}) it is clear that the factor $\ell_p$ for
$P\neq \theta$ is given by
\begin{eqnarray} \ell_p &=& - e^{-K}\(A_{I J}\oline{A}^{I J}\)_p\wh{V} +
e^{-K}\[A_{I J k}\oline{A}^{I J}_{\bar{m}} + \(K_{i\bar{m}}A_{I
J}\oline{A}^{I J} + 2R^L_{k\bar{m} I}A_{L J}\oline{A}^{I
J}\)\]_p\(\D^{\mu} z^k\D_{\mu}\bar{z}^{\m} +
F^k\oline{F}^{\bar{m}}\) \nonumber \\
 & & - \lbr e^{-K}(A_{I J k}\oline{A}^{I J})_p\[F^k\mG + \frac{1}{2x}
 \D_a(T^a z)^k\]+\hc\rbr, \end{eqnarray}
which is the same as the first 3~lines of~(\ref{everything}) with
\begin{equation} \wh{V} = K_{i\bar{m}}F^i\oline{F}^{\bar{m}}
- 3\mG^2, \qquad F^i = e^{-K/2}\oline{A}^i, \qquad \mG^2 =
e^{-K}A\oline{A}. \label{defs}\end{equation}
%

The same calculation for $P = \theta$ is only slightly more
complicated. In this case
\begin{eqnarray} \pp_k\delta k^b &=& \delta k^b_k = \(\alpha_b^{\theta} K_k +
\pp_k\ln\mu_\theta\)\delta k^b = \Gamma^b_{b k}\delta k^b,
\nonumber \\
\pp_{\bar{m}}\pp_k\delta k^b &=& \delta k^b_{k\bar{m}} =
\[\(\alpha_b^\theta K_k + \pp_k\ln\mu_\theta\) \(\alpha_b^\theta
K_{\bar{m}} + \pp_{\bar{m}}\ln\bar\mu_\theta\) + \alpha_b^\theta
K_{k\bar{m}}\]\delta k^b \nonumber \\
 &=& \(R^b_{b k\bar{m}} + \Gamma^b_{b k}\Gamma^{\bar b}_{\bar b\bar{m}}\)
 \delta k^b ,
\label{ktheta}\end{eqnarray}
so that the corresponding $\ell_b$ is given by
\begin{eqnarray} \ell_b/\delta k^b &=& - \wh{V} + R^b_{b k\bar{m}}
\(\D^\mu z^k\D_\mu\bar{z}^{\bar{m}} + F^k\oline{F}^{\bar{m}}\) +
\Gamma^b_{b k}\Gamma^{\bar b}_{\bar b\bar{m}}\(\D^\mu
z^k\D_\mu\bar{z}^{\bar{m}} + F^k \oline{F}^{\bar{m}}\) \nonumber
\\
& & - \lbr \Gamma^b_{b k}\[F^k\mG + \frac{1}{2}\D_a(T^a
z)^k\]+\hc\rbr .
\label{ellb}\end{eqnarray}
From~(\ref{lambdas}) and the definitions in~(\ref{matrix}) we have
\begin{equation}
\delta k^b =  - \frac{4}{x_b}K_{b\bar b}|\delta_b\theta^b|^2
\end{equation}
and after some algebra the entries in~(\ref{ellb}) can be
identified with the final three lines of~(\ref{everything}).

Finally, as emphasized above, the effective one loop action given
in~(\ref{Lagrenorm}) respects supersymmetry only to lowest order
in the loop expansion parameter $\epsilon = \hbar/16\pi^2$. More
precisely if $g_{\mu\nu}^R$ and $K^R = K + \delta K$ are the fully
renormalized metric and K\"ahler potential, then~(\ref{Lagrenorm})
can be written as
\begin{equation}
\Lag_{\rm tree}(g^R,K^R) =  \Lag_{\rm tree}(g,K) + \Lag_{\rm
1-loop} + O(\epsilon\ln\Lambda^2_{\rm eff}) + O(\epsilon^2),
\end{equation}
where the leading $N_{\chi}$ and $N_G$ corrections that give
potentially significant contributions to the effective potential
at one loop are
\begin{equation}
\delta K = \frac{1}{32\pi^2}\[N_{\chi}\Lambda_{\chi}^2 -
4N_G\Lambda_G^2 +O(1)\Lambda^2_{\rm grav}\] =
\frac{\epsilon}{2}\[N_{\chi}\Lambda_{\chi}^2 - 4N_G\Lambda_G^2
+O(1)\Lambda^2_{\rm grav}\] .
\label{delk} \end{equation}
If the coefficients of the various $\Lambda^2$ terms are of order
unity then in order to preserve local supersymmetry of the
Lagrangian we have to retain all terms that follow from the
correction~(\ref{delk}) to the K\"ahler potential; this amounts to
summing leading terms in $(\epsilon N\Lambda^2_\chi)^n$ and
$(\epsilon N_G\Lambda^2_G)^n$, with the result of the sum just
giving a correction to the K\"ahler potential as dictated by local
supersymmetry.

\newpage
\section{Phenomenological Applications in String-Based Models}
\label{sec:phenom}

In the preceding sections we demonstrated how the leading one-loop
correction to the scalar potential, which is quadratic in the
cutoff, is obtained. The explicit form of this correction was
given in~(\ref{Hlight2}) and shown to be of the general form
of~(\ref{STr}). We then added a supersymmetric regulating sector
to the theory in the form of a set of massive Pauli-Villars
regulator fields. These add new contributions to the one-loop
effective potential, rendering the divergent contributions finite.
Formally divergent contributions to the one-loop Lagrangian are
now replaced with terms proportional to the square of various
field-dependent mass terms. The combined contribution is given
in~(\ref{everything}), providing an explicit realization of the
subtractions mentioned in~(\ref{subtract}).

While it is possible to work directly with the (regularized)
Lagrangian represented by~(\ref{everything}), doing so ignores the
great power of working with a supersymmetric regulating sector --
namely, the ability to treat the quadratically divergent
corrections as a renormalization of the K\"ahler potential itself.
This correspondence, as written in~(\ref{Lexpand}), was
demonstrated in Section~\ref{sec:renorm}. In this section we will
take the renormalized K\"ahler potential of~(\ref{tildek}) as our
starting point in looking at the phenomenological implications of
these corrections on vacuum stability and scalar mass terms. For
most of this section we will be interested primarily in those
terms that involve fields connected with the gauge-charged chiral
matter of the light spectrum. The relevant corrections to the
K\"ahler potential are therefore those of~(\ref{ddelk}), and we
will make use of the relations in~(\ref{daij}).

\subsection{Vacuum Energy and Scalar Masses}
\label{sec:vac} As a first example, suppose that the effective cut-offs are
field-independent constants. In this case the quantum corrected
effective potential is just
\begin{equation} V_{eff} = \D + e^{K+\Delta K}\(F^i K_{i\bar{m}}
\oline{F}^{\bar{m}} - \frac{1}{3}|M|^2\)_{\rm tree} ,
\label{const}\end{equation}
where $M$ is the auxiliary field of supergravity whose vacuum
expectation value determines the gravitino mass through the
equation of motion $M = -3e^{K/2}W$. In particular, if
supersymmetry breaking is F-term induced, {\em i.e.}
$\lang\D\rang=0$, then the tree level condition $F^i K_{i\bar{m}}
\oline{F}^{\bar{m}} = \frac{1}{3}|M|^2$ for vanishing vacuum
energy is unmodified by these quantum corrections. In other words,
even after including all quadratically divergent contributions to
the one-loop effective potential (with a proper supersymmetric
regularization), the vacuum energy will continue to vanish
provided $\lang V_{\rm tree} \rang =0$ is satisfied.

In some circumstances this simplification can be achieved. For
example it was shown in~\cite{Gaillard:1999ir} that the masses of
PV fields that regulate the (logarithmically divergent) untwisted
matter loop corrections involving renormalizable couplings can be
made constant (and thereby modular invariant). However not all
masses can be constant and modular invariant. Regulation of
quadratic and logarithmic divergences requires (for example) some
masses proportional to $e^K$ in the matter + gravity sector and
proportional to $g^2_{\STR} =(s + \bar{s})/2$ in the gauge +
dilaton sector. Modular invariance of the one-loop corrected
Lagrangian then requires either that terms proportional to $e^K$
cancel in the supertrace implicit in~(\ref{tildek}), or that the
mass terms in the PV~superpotential have a T-dependence that
restores modular invariance; the latter would be interpreted as a
parameterization of string loop threshold
corrections.\footnote{There is a residual noninvariance associated
with terms logarithmic in the PV masses that is canceled by the
Green-Schwarz term.} Thus we generally expect some modification of
the effective potential for moduli in string-based effective
supergravity theories, as was anticipated by~\cite{Choi:1994xg}.

For concreteness, in the remainder of this section we consider the
BGW model~\cite{Binetruy:1996nx,Binetruy:1997vr} for gaugino
condensation, in which dilaton stabilization was
achieved~\cite{Binetruy:1996xj,Casas:1996zi} by invoking
nonperturbative string~\cite{Shenker:1990uf,Silverstein:1996xp}
and/or QFT~\cite{Banks:1994sg,Giedt:2003ap} corrections to the
dilaton K\"ahler potential.  In order to implement the correct
Bianchi identity for the gaugino condensate composite superfield
-- as well as the Green-Schwarz (GS) anomaly cancellation
mechanism -- it is much more convenient to use the linear
multiplet formulation for the dilaton, as was done
in~\cite{Binetruy:1996nx,Binetruy:1997vr}.
In~\cite{Binetruy:2000md} the results were recast in the more
familiar language of the chiral multiplet formalism, with the
effective tree level potential below the scale of condensation
taking the standard form $V=F^iK_{i\bar{m}}\oline{F}^{\bar{m}} -
\frac{1}{3}M\oline{M}$ with
\beq M = \half b_c u, \qquad F^S = - {1\over4}K_{S\S}^{-1}\(1 -
{2\over3}b_c K_S\)\bar u, \eeq
where $u$ is the vacuum value of the gaugino condensate. The
quantity $b_c$ is the beta function coefficient of the condensing
gauge group ${\cal{G}}_c$:
\beq b_c = {1\over16\pi^2}\(3C_c - \sum_i C_c^i\),\eeq
where $C_c$ and $C^i_c$ are the quadratic Casimirs in the adjoint
and matter representations, respectively, of ${\cal{G}}_c$.  The
BGW model is explicitly modular invariant and the K\"ahler moduli
$T^i$ are stabilized at self-dual points with vanishing auxiliary
fields: $\lang F^{T^i}\rang =0$. Supersymmetry breaking is
therefore dilaton-dominated and the condition for vanishing vacuum
energy at tree level in the effective theory is
\beq \lang V_{\rm eff}\rang = \lang K_{S\S}|F^S|^2 -
{1\over3}|M|^2\rang = 0, \qquad \lang K^{-1}_{S\S}\rang =
{4b_c^2\over3(1 - {2\over3}\lang K_S\rang b_c)^2} .
\label{vacen}\eeq
Classically,
\beq -2\lang K_S\rang = 2\lang K^{\half}_{S\S}\rang = 2\lang(s +
\s)^{-1}\rang = g_{\STR}^2 \approx \half , \label{gstr} \eeq
where $g_{\STR}$ is the value of the gauge coupling constant at
the string scale. The last approximate relation in~(\ref{gstr})
can be inferred from low energy data extrapolated to high energy
scales through renormalization group (RG) evolution.

The model is both phenomenologically~\cite{Gaillard:1999et} and
cosmologically~\cite{Birkedal-Hansen:2001is,Birkedal-Hansen:2002am}
viable if $b_c\approx .05$--$.06$, so it is clear
that~(\ref{vacen}) cannot be satisfied without invoking
nonperturbative string effects that modify the K\"ahler potential
for the dilaton. The suppression of the vacuum value $\lang
K_{S\S}^{-1}\rang$ with respect to its classical value implied
by~(\ref{vacen}) has the effect of enhancing the dilaton mass,
which is a welcome feature for modular cosmology. However it also
entails a suppression of gaugino masses relative to scalar masses,
which increases somewhat the fine tuning problem of MSSM
phenomenology~\cite{Kane:2002ap}. The problem is
exacerbated~\cite{Barreiro:1998nd,Gaillard:2003gt} when a D-term
arising from an anomalous $U(1)$ is included. Except in a class of
models in which the preferred vacuum configuration minimizes the
number of large $vev$s $\lang\phi^a\rang$, a further suppression
of the dilaton F-term is required to maintain vanishing vacuum
energy at tree level. In this case a large, positive constant
value of $\Delta K$ in~(\ref{const}) would be welcome. In addition
there are dangerously large and potentially negative D-term
contributions to squark and slepton masses unless $-K_S$ is
considerably larger than its classical value, thereby calling into
question the validity of the weak coupling approximation for these
models~\cite{Banks:1994sg,Witten:1996mz}.

Now let us consider the possible impact on the dilaton potential
of field dependent effective cut-offs of the type discussed above.
With the Green-Schwarz term included modular invariance dictates a
renormalized K\"ahler potential of the form
\beq K^R = K + \half\epsilon c_\chi f(T,\T)e^K - {4\epsilon
c_G\over S + \S - V_{\GS}} = K + \half\epsilon c_\chi f(T,\T)e^K -
2\epsilon c_G g_s^2(Z,\bar Z),\label{newk}\eeq
where $V_{\GS}(T + \T)$ is the Green-Schwarz term and $f(T,\T)$ is
a function with a well-defined transformation property under
modular symmetries that assures modular invariance of the second
term. The loop-induced terms proportional to $\epsilon$ may not be
negligible if $c_\chi,c_G\sim N_\chi,N_G$, respectively. Since the
theory is still modular invariant we expect that the moduli are
still stabilized at self-dual points, where the additional
contributions to $F^{T^i}$ induced by these quantum corrections
vanish. Setting the T-moduli at their $vev$s, defining
\beq \tc_\chi = \lang{f(t,\t)e^G}\rang c_\chi,\eeq
and setting the matter fields to zero, the renormalized potential
and its $S$-derivatives read
\bea K^R &=& k + {\epsilon\over2}\(\tc_\chi e^k - 4c_G
g^2_{\STR}\),\qquad k = k(2g_{\STR}^{-2}),\qquad K^R_S = K_S\(1 +
{\epsilon\over2}\tc_\chi e^k\) + \epsilon c_G g_{\STR}^4,\nonumber
\\ K^R_{S\S} &=& K_{S\S}\(1 + {\epsilon\over2}\tc_\chi e^k\) +
{\epsilon\over2}\(K^2_S\tc_\chi e^k - c_G g_{\STR}^8\).\eea
The condition for vanishing vacuum energy is still given
by~(\ref{vacen}) but with the replacement $K \to K^R$. The
relevant parameter for particle physics phenomenology is now the
\vev\, of $1/K_{S\S}^R$, which remains strongly suppressed with
respect to its classical value since $K^R_S$ is negative
semi-definite.\footnote{The relation $\ell = - K^R_S$ holds at any
given order in perturbation theory, where $\ell$ is the dilaton of
the dual linear multiplet formalism.} Therefore the salient
phenomenological features of the BGW model are essentially
unaffected by these quantum corrections.

On the other hand, could these corrections lessen the need to
invoke rather large nonperturbative effects, especially when an
anomalous $U(1)$ is incorporated?  A large {\it negative} value of
$c_G$ or a large {\it positive} value of $\tc_\chi$ would increase
$-K^R_S$ and decrease $1/K^R_{S\S}$ for fixed $g^2_{\STR} \approx
1/2$, which is the desired effect.  One can reasonably assume that
$|c_G|\le N_G \le 65 \sim 0.4\epsilon^{-1}$ in typical orbifold
compactifications~\cite{Giedt:2001zw}, so a significant effect
cannot be obtained from the second term in~(\ref{newk}).  On the
other hand $N_{\chi} \epsilon\sim 2$ for typical orbifolds. Quite
generally we have
\beq  f(t,\t) = \prod_i[|\eta(t^i)|^4(t^i + \t^i)]^{q_i},\qquad
\lang|\eta(t^i)|^4(t^i + \t^i)\rang\approx 1,\eeq
where the last approximate equality holds at a self-dual point. In
this case if $\tc_\chi \sim N_{\chi}$ and $e^k\sim 1$, it might be
possible to somewhat alleviate the above-mentioned difficulties of
condensation models.

One would also expect corrections to the K\"ahler potential to
affect the overall size of scalar masses (we will address the
off-diagonal structure of these masses in the next subsection). In
the condensation models (with or without an anomalous $U(1)$)
considered here the scalar mass terms in the effective potential
arise from derivatives of the K\"ahler potential, and the K\"ahler
metric factors out of the normalized squared mass if the K\"ahler
potential is of the form
\beq K = k(S,\S) + g(T,\T) + \sum_a\kappa_a(T,\T)|\Phi^a|^2 +
O(\Phi^3).\label{sep}\eeq
If the second term in~(\ref{newk}) is present, the K\"ahler
potential is no longer separable as in~(\ref{sep}), and the F-term
in the potential takes the form (again setting the moduli at their
self-dual points)
\beq V_F = K^R_{S\S}|F^S|^2 + K^R_{a\bar b}F^a\bar F^{\bar b} +
\(K^R_{S\bar b}F^S\bar F^{\bar b} + {\rm h.c.}\), \qquad F^a =
 {b_c\bar u\over6}K_R^{a\bar b}K^R_{\bar b} .\eeq
The wave function normalization factor $\kappa_a/(1 +
\half\epsilon\tc e^k)$ again cancels, and -- neglecting the third
term in~(\ref{newk}) -- we obtain for the squared mass of the
canonically normalized gauge-charged scalar~$\phi^a$
\bea m^2_{\phi^a} &=& {b_c^2|u|^2\over36}\[1 +
{3\over2}\epsilon\tc e^k{1-{2\over3}b_c K^R_S\over
K^R_{S\S}}\({1-{2\over3}b_c K^R_S\over K^R_{S\S}} - b_c K^R_S\)\]
+ {1\over s}\sum_b q^b_a D_b \nonumber\\ &=&
{b_c^2|u|^2\over36}\[1 + {2b_c^3\epsilon\tc e^k\over1-{2\over3}b_c
K^R_S}\({4b_c\over3(1-{2\over3}b_c K^R_S)} - K^R_S\)\] + {1\over
s}\sum_b q^b_a D_b, \eea
where in the last expression we used the vacuum
condition~(\ref{vacen}). Even if $\tc_\chi\sim1$, these
corrections are subleading if $b_c\ll1$.

The primary conclusion of this section is that the only practical
effect of the leading $N_\chi,N_G$ quadratic divergences  in this
particular concrete example is a possible re-interpretation of the
modification of the dilaton superpotential, needed for
stabilization, in terms of a combination of string nonperturbative
effects and quantum field theory perturbative effects, as opposed
to only the former. We note the importance of the supersymmetric
regularization in allowing an interpretation of the quadratic
contributions to the dilaton effective potential in terms of
renormalization of the K\"ahler potential as in~(\ref{newk}). This
interpretation made it possible to quickly see the muted impact of
these terms, even in the case of large $c_G$ and $\tc_{\chi}$.

\subsection{Potential Off-Diagonal Scalar Mass Terms}
\label{sec:offdiag} To address the question of what constraints are needed to avoid
experimentally excluded flavor changing neutral current (FCNC)
effects, we first note that the tree potential of an effective
supergravity theory includes a term
\beq V_{\rm tree}\ni e^K K_i
K_{\jbar}K^{i\jbar}|W|^2.\label{vkahl}\eeq
So prior to any discussion of large loop-induced contributions to
flavor-changing operators it is necessary to ensure their absence
at the tree level. The observed suppression of FCNC effects thus
constrains the K\"ahler potential already at the leading order --
to a high degree of accuracy we require that
\beq K_i K_{\jbar}K^{i\jbar} \not{\hspace{-.03in}\ni} \lang
f(X,\bar X)\rang \phi^a_f\bar\phi^{\bar a}_{f'\ne f},\eeq
where $f,f'$ are flavor indices, $a$ is a gauge index, $\phi^a_f$
any standard model squark or slepton, and $X$ is a singlet of the
Standard Model gauge group. For example, in the no-scale models
that characterize the untwisted sector of orbifold
compactifications, we have
\beq K_i K_{\jbar}K^{i\jbar} = 3 + K_S K_{\bar S} K^{S\bar S},\eeq
which is safe, since $K_S$ is a function only of the dilaton.  The
twisted sector K\"ahler potential is known only to quadratic
order:
\beq K_T = \sum_a e^{g^a(T + \bar T)}|\Phi^a_T|^2 + O(\Phi^3),
\label{tw}\eeq
which is flavor diagonal and also safe.  The higher order terms in
(\ref{tw}) could be problematic if some $\phi^a = X^a$ have large
$vev$s ({\it i.e.} within a few orders of magnitude of the Planck
scale).  Thus phenomenology requires that we forbid couplings of
the form $\phi^a_f\phi^{\bar a}_{f'\ne f} |\phi^{a'}_{f"}|^2
X^{b_1}\cdots X^{b_n}$, $n\le N$, where $N$ is chosen sufficiently
large to make the contribution $\lang X^{b_1}\cdots X^{b_n}\rang$
to the scalar mass matrix negligible.

The quadratically divergent one-loop corrections generate a term
\beq V_{\rm 1-loop}\ni e^K K_i K_{\jbar}R^{i\jbar}|W|^2, \qquad
R^{i\jbar} = K^{i\bar k}R_{\bar k l} K^{k\jbar}.\label{vricc}\eeq
where $R_{i\jbar}$ is the K\"ahler Ricci tensor.  The
contribution~(\ref{vricc}) simply reflects the fact that the
leading divergent contribution in a nonlinear sigma model is a
correction to the K\"ahler metric proportional to the Ricci tensor
(whence {\it e.g.} the requisite Ricci flatness of two dimensional
conformal field theories).  Since the Ricci tensor involves a sum
of K\"ahler Riemann tensor elements over all chiral degrees of
freedom, a large, order $N_\chi$, coefficient may be generated.
For example, for an untwisted sector $U$ with three untwisted
moduli $T^i$ and K\"ahler potential
\beq K^U = \sum_{n=1}^3K^n = - \sum_{n=1}^3\ln(T^n + \bar T^{\bar
n} - \sum_{a=1}^{N_n}|\Phi^a_n|^2),\label{nskp} \eeq
we get
\beq R^n_{i\jbar} = (N_n + 2)K^n_{i\jbar}.\label{runtw}\eeq
While this contribution is clearly safe, since the Ricci tensor is
proportional to the K\"ahler metric, the condition that the tree
potential be FCNC safe does not by itself ensure
that~(\ref{vricc}) is safe in general.  For this we require in
addition the absence of K\"ahler potential terms of the form
$\phi^a_f\bar\phi_{f'\ne f}^{\bar a}|\phi_{f''}^{a'}|^4(X^b)^{n\le
N}$.  On the other hand, if the K\"ahler metric is FCNC safe due
to an {\it isometry}, the same isometry will protect the Ricci
tensor from generating FCNC.

For example, the scalar metric $g_{i j}$ for the effective pion
Lagrangian is dictated by chiral $SU(2)_L\times SU(2)_R$; there is
a unique form of the two-derivative coupling:
\beq g_{i j} = \delta_{i j} + {\pi_i\pi_j\over v^2 - \pi^2},\eeq
for a particular choice of field variables.  Preservation of this
symmetry at the one-loop level assures that $R_{i j}\propto g_{i
j}$. Similarly, the kinetic term derived from the K\"ahler
potential~(\ref{nskp}) possesses an $\prod_{n=1}^{3} SU(N_n +
1,1)$ symmetry that is much larger than the $SL(2,R)$ (or
possibly$[SL(2,R)]^3$) T-duality symmetry of the full Lagrangian,
and we obtain the result~(\ref{runtw}).  More generally, in
effective supergravity from string compactifications there are a
number of selection rules and/or symmetries that forbid
superpotential couplings that are allowed by gauge invariance and
the T-duality invariance group (see for
example~\cite{Font:1989aj}).  The K\"ahler potential has not been
investigated in similar detail, but {\it a priori} one would
expect an analogous pattern. In the absence of input from string
theory one can work backwards and ask: what constraints does
phenomenology impose?

We first note that there is a large class of models in which FCNC
are suppressed independently of the details of the structure of
the K\"ahler potential, provided the moduli $t^I$ are stablized at
self dual points. The supersymmetric completion of the potential
in any given order in perturbation theory yields (in the absence
of D-term contributions) the scalar squared mass matrix
\beq (m^2)^i_j = \delta^i_j\mG^2 - \lang \wtd{R}^i_{j
k\bar{m}}\rang \wtd{F}^k \wtd{\oline{F}}^{\bar{m}}, \label{mtree}
\eeq
where $\wtd{R}^i_{j k\bar{m}}$ is an element of the Riemann tensor
derived from the fully renormalized K\"ahler metric, and
$\wtd{F^i}$ is the auxiliary field for the chiral superfield
$\Phi^i$, evaluated by its equation of motion using the quantum
corrected Lagrangian. Since the latter is perturbatively modular
invariant, the K\"ahler moduli $t^i$ are still stabilized at
self-dual points with $\lang \wtd{F}^{t^i}\rang=0$. Classically we
have $R^a_{b s\bar{s}}=0$ where the indices $a,b$ refer to
gauge-charged fields in the observable sector. This need not be
true at the quantum level. For example, if, as suggested
in~(\ref{newk}), the quantum correction to the K\"ahler potential
includes a term
\beq \Delta K = \frac{1}{32\pi^2}\STr\Lambda^2_{\rm eff} \ni
\frac{c N_{\chi}}{32\pi^2}e^{\alpha K},\eeq
we get
\beq \lang \wtd{R}^a_{b s\bar{s}} \rang = \delta^a_b\frac{c
N_{\chi}}{32\pi^2} \alpha^2 e^{\alpha K}\(K_{s\bar{s}} + \alpha
K_s K_{\bar{s}}\),\eeq
which is flavor diagonal, and therefore FCNC safe.

To consider more general situations, let us write out the elements
of the renormalized K\"ahler Riemann tensor
\begin{eqnarray}
\wtd{R}^i_{j k\bar{m}} &=& \wtd{D}_{\bar{m}}\wtd{\Gamma}^i_{k j} =
\wtd{K}^{i\bar{n}}\wtd{D}_{\bar{m}}\wtd{\Gamma}_{\bar{n} k j} =
\wtd{K}^{i\bar{n}}\[\pp_{\bar{m}}\wtd{\Gamma}_{\bar{n} k j} -
\wtd{\Gamma}^{\bar{r}}_{\bar{m}\bar{n}}\wtd{\Gamma}_{\bar{r} k
j}\] \nonumber \\
 &=& K^{i\bar{n}}\[D_{\bar{m}}\wtd{\Gamma}_{\bar{n} k j} +
 \(\Gamma^{\bar{r}}_{\bar{m}\bar{n}} -
\wtd{\Gamma}^{\bar{r}}_{\bar{m}\bar{n}}\)\wtd{\Gamma}_{\bar{r} k
j}\], \nonumber \\
\wtd{\Gamma}_{\bar{n} k j} &=& \pp_k\wtd{K}_{\bar{n} j} =
\Gamma_{\bar{n} k j} + \pp_i \delta K_{\bar{n} j} =
\Gamma_{\bar{n} k j} + D_i \delta K_{\bar{n} j} + \Gamma^l_{k j}
\delta K_{\bar{n} l} ,
\label{curve}\end{eqnarray}
with the inverse metric $\wtd{K}^{i\bar{n}}$ given by
\begin{equation} \wtd{K}^{i\bar{n}} = K^{i\bar{n}} - K^{i\bar{p}}
\delta K_{\bar{p} l}K^{l\bar{n}} + O(\delta K^2). \end{equation}
The potentially dangerous terms in~(\ref{mtree}) can then be
extracted from consideration of~(\ref{curve}), for which we have
\beq \wtd{R}^i_{j k\bar{m}} = R^i_{j k\bar{m}} -
K^{i\bar{n}}\delta K_{\bar{n} l}R^l_{j k\bar{m}} +
K^{i\bar{n}}D_{\bar{m}}D_k\delta K_{\bar{n} j} + O(\delta K^2),
\eeq
and the auxiliary fields for the chiral superfields, given by
\beq \wtd{F}^i = e^{\wtd{K}/2}\wtd{K}^{i\jbar}\(\oline{W}_{\jbar}
+ \oline{W}\wtd{K}_{\jbar}\) = F^i\(1 + \frac{\delta K}{2}\) -
K^{i\jbar} \delta K_{\jbar k}F^k +  e^{K/2}K^{i\jbar}\delta
K_{\jbar}\oline{W} + O(\delta K^2).\eeq
Then the second term in~(\ref{mtree}) is given by
\begin{eqnarray}
(m^2_R)^i_j &=& - \wtd{R}^i_{j k\bar{m}} \wtd{F}^k
\wtd{\oline{F}}^{\bar{m}} \nonumber \\
 &=& - R^i_{j k\bar{m}}F^k \oline{F}^{\bar{m}}\(1 + \frac{\delta K}{2}\) +
R^{i\;\bar{m}}_{\;j\;\;\bar{n}}F^l \oline{F}^{\bar{n}}\delta
K_{l\bar{m}} + R^{i\;\;\;l}_{\;j k}F^k \oline{F}^{\bar{m}}\delta
K_{l\bar{m}} \nonumber \\
 & & - e^{K/2}\(R^{i\;\bar{m}}_{\;j\;\;\bar{n}} \oline{F}^{\bar{n}}
 \delta K_{\bar{m}}\oline{W} + R^{i\;\;\;l}_{\;j k}F^k \delta K_l
W\) \nonumber \\
 & & + R^l_{j k\bar{m}}F^k \oline{F}^{\bar{m}}K^{i\bar{n}} \delta
 K_{l\bar{n}} - K^{i\bar{n}}F^k \oline{F}^{\bar{m}} D_{\bar{m}}D_j
 \delta K_{k\bar{n}} + O(\delta K^2). \label{m2}\end{eqnarray}

Let us investigate the consequences for the following simple
K\"ahler potential
\begin{equation} K = g(M,\oline{M}) + \sum_a f_a(M,\oline{M})|\Phi^a|^2 +
O(\Phi^3), \label{defwt} \end{equation}
where $M$, $\oline{M}$ represent chiral superfields of the hidden
sector (such as moduli in a string construction) which are SM
gauge singlets, and we have chosen a basis in which the term
quadratic in the matter fields $\Phi^a$ is diagonal. The
tree-level contribution to the scalar masses is then
\begin{equation} \lang R^a_{b n\bar{m}}F^n
\oline{F}^{\bar{m}}\rang = \delta^a_b \lang g^a_{n\bar{m}}
F^n\oline{F}^{\bar{m}}\rang, \qquad g^a_{n\bar{m}} =
f_a^{-1}\pp_{\bar{m}} \pp_n f_a - f_a^{-2}\pp_{\bar{m}}f_a\pp_n
f_a. \end{equation}
To avoid tree-level FCNC, we require $\lang g^a_{n\bar{m}}
F^n\oline{F}^{\bar{m}}\rang$ to be independent of flavor for fixed
SM gauge quantum numbers. As a specific example, in string theory
with $f_a = (T^i + \oline{T}^i)^{-q_a^i}$ we can avoid tree-level
FCNC if $\lang F^{t^i} \rang =0$ or if all quarks with the same
quantum numbers have the same modular weights\footnote{Note that two
independent conventions for modular weights exist in the literature.
In particular, if we use the expression in~(\ref{defwt}) with $f_a =
(T^i + \oline{T}^i)^{-q_a^i}$ to serve as the defining property of
the integer modular weight, then our convention corresponds to those
of~\cite{Gaillard:1992bt,Binetruy:1996nx,Binetruy:1997vr,Binetruy:1996xj},
while the case $-q_a^i \to n_a^i$ is the convention
of~\cite{Binetruy:2000md,Ibanez:1992hc,Kane:2002qp,Kane:2004tk}. The
sign convention on the weights we have chosen has the virtue that
the integers $q_a^i$ will typically be positive.} $q_a^i$:
$g^a_{i\jbar} = \delta_{ij} \, q^i_a(2\re t^i)^{- 2}$. Assuming tree
level FCNC are absent, the only dangerous part of~(\ref{m2}) is the
last line. Using~(\ref{daij}),~(\ref{ddelk}), and
\begin{eqnarray} D_l\(e^{-K}\oline{A}^{I J}_{\bar{m}}\) &=&
e^{-K}\(K_{l\bar{m}}\oline{A}^{I J} R_{P\bar{m} l}^{I}
\oline{A}^{PJ} + R_{P\bar{m} l}^{J}\oline{A}^{IP}\), \nonumber \\
D_l\delta k_{k\bar{m}} &=& K_{k\bar{m}}\delta k_l +
K_{l\bar{m}}\delta k_k \nonumber \\
 & & + e^{-K}\[A_{I J k
l}\oline{A}^{I J}_{\bar{m}} + 2\(R_{I\bar{m} l}^{P} A_{P J
k}\oline{A}^{I J} + [k\lra l]\) + 2\(D_l R_{I\bar{m} k}^{P}\) A_{P
J}\oline{A}^{I J}\]\nonumber \\
D_{\bar{m}}\(e^{-K}A_{I J k l}\) &=& e^{-K}\lbr R^n_{l\bar{m}
k}A_{I J n} +
\[K_{k\bar{m}}A_{I J l} + \(R^P_{I\bar{m} k}A_{P J l} +[I\lra J]\) + (k\lra
l)\]\rbr \nonumber \\
 & & + e^{-K}\[A_{P J}D_l R^P_{I\bar{m} k} + (I\lra
 J)\] \end{eqnarray}
we obtain for the modes $P\ne\theta$
\begin{eqnarray}
D_{\bar{n}}D_l\delta k_{k\bar{m}} &=& e^{-K}A_{I J k
l}\oline{A}^{I J}_{\bar{m}\bar{n}} + \[K_{k\bar{m}}\delta
k_{l\bar{n}} + (k\lra l) + (n\lra m)\] -
\[K_{k\bar{m}}K_{l\bar{n}} + (k\lra l)\]\delta k \nonumber \\
& & + e^{-K}\[R^p_{l\bar{n} k}A_{I J p}\oline{A}^{I J}_{\bar{m}} +
2\(D_{\bar{n}}D_l R^P_{I\bar{m} k}\)A_{P J}\oline{A}^{I J}\]
\nonumber \\
 & & + 2e^{-K}\lbr\[\(D_l R^P_{I\bar{m} k}\)A_{P J}\oline{A}^{I J}_{\bar{n}} +
(n\lra m)\] +
\[\(D_{\bar{n}} R^P_{I\bar{m} l}\)A_{P J k}\oline{A}^{I J} + (k\lra
l)\]\rbr \nonumber \\
& & + 2e^{-K}\[R^P_{I\bar{m} l}A_{P J k} \oline{A}^{I J}_{\bar{n}}
+ (n\lra m) + (k\lra l)\] \nonumber \\
& & + 2e^{-K}\[R^Q_{I\bar{m} l}R^P_{Q\bar{n} k}A_{P J}\oline{A}^{I
J} + R^Q_{I\bar{m} l}R^P_{J\bar{n} k}A_{P Q}\oline{A}^{I J} +
(k\lra l)\]. \label{dddelk} \end{eqnarray}

Inserting this into the last line of~(\ref{m2}) gives
\begin{eqnarray} (m^2_P)^a_b &\ni& - F^m\oline{F}^{\bar{n}}\lbr\delta_b^a\delta k_{m\bar{n}}
+ K_{m\bar{n}}K^{a\bar{c}}\delta k_{b\bar{c}} - \delta_b^a
K_{m\bar{n}}\delta k - R^c_{bm\bar{n}}K_{a\bar{d}} \delta k_{c\bar{d}}
\right. \nonumber \\
& & \quad + 2e^{-K}\[\(D_{\bar{n}}D_b R^{P\;a}_{\;I\;\; m}\)A_{P
J}\oline{A}^{I J} + \(D_b R^{P\;a}_{\;I\;\; m}\)A_{P
J}\oline{A}^{I J}_{\bar{n}} + 2\(D_{\bar{n}} R^{P\;a}_{\;I\;\;
b}\)A_{P J m}\oline{A}^{I J}\right. \nonumber \\
 & & \quad \left. \left.
+ R^{P\;a}_{\;I\;\; b}A_{P J m} \oline{A}^{I J}_{\bar{n}} +
R^{Q\;a}_{\;I\;\; b}R^P_{Q\bar{n} m}A_{P J}\oline{A}^{I J} +
R^{Q\;a}_{\;I\;\; b}R^P_{J\bar{n} m}A_{P Q}\oline{A}^{I J}\]\rbr,
\label{m3}\end{eqnarray}
where we dropped terms that vanish in the vacuum, and the above
``squared masses'' have to be put in a weighted sum as
in~(\ref{tildek}) with the appropriate loop factor.

Most of the terms in~(\ref{m3}) are in fact proportional to
$\delta_a^b$, so represent corrections to the diagonal elements of
the scalar mass matrix. The off-diagonal elements are the result
of terms involving the set of PV fields $P= \Phi^I$, in which we
identify $R^I_{J k l}\to R^i_{j k l}$:
\begin{eqnarray}
(m^2_{\Phi^I_\alpha,\Pi^I_\alpha})^a_b &\ni& - e^{-K}F^m
\oline{F}^{\bar{n}}\[ K_{m\bar{n}}R^{p\;a}_{\;i\;\; b}A_{P
J}\oline{A}^{I J} + \(D_{\bar{n}}D_b
R^{p\;a}_{\;i\;\; m}\)A_{P J}\oline{A}^{I J}\right. \nonumber \\
 & & \left. + \(D_b R^{p\;a}_{\;i\;\; m}\)A_{P J}\oline{A}^{I J}_{\bar{n}}
 + \(D_{\bar{n}} R^{p\;a}_{\;i\;\; b}\)A_{P J m}\oline{A}^{I J} \right. \nonumber \\
 & & \left. +
R^{p\;a}_{\;i\;\; b}A_{P J m} \oline{A}^{I J}_{\bar{n}} - \alpha
R^{p\;a}_{\;i\;\; b}K_{\bar{n} m}A_{P J}\oline{A}^{I J} \right.
\nonumber \\
 & & \left. - R^a_{b m\bar{n}}A_{I J}A^{I J} - 2R^c_{b m\bar{n}}R^{P\;a}_{\;I\;\;
c}A_{P J}A^{I J} \] .
\label{m4} \end{eqnarray}
Using the quantities in~(\ref{matrix}) and~(\ref{connections}),
along with the identity
\begin{equation}
2R^P_{I\bar{m} k}A_{P J}\oline{A}^{I J} =  \(R^\Phi_{k\bar{m}} +
R^\Pi_{k\bar{m}}\) |\beta|^2\Lambda^2,
\end{equation}
the potentially dangerous mass terms in~(\ref{m4}) can be written
as\footnote{In what follows we have a dropped a term $\beta^2
\Lambda^2 R^a_{bm\bar{n}}F^m \oline{F}^{\bar{n}}$ since it will not
produce any FCNC contributions if the tree-level masses are FCNC
safe.}
\begin{eqnarray}
(m^2_{\Phi^I_\alpha,\Pi^I_\alpha})^a_b &\ni& K^{a\bar{c}}F^m
\oline{F}^{\bar{n}}\beta^2_\alpha\Lambda^2_\alpha \lbr
D_{\bar{n}}D_b R_{m\bar{c}} +2R_{d\bar{c}} R^d_{bm\bar{n}}\right. \nonumber \\
 & & \left.+ \(
K_{m\bar{n}}\(1 -\alpha_\alpha\) + \[(1 - \alpha_\alpha)K_m -
\pp_m\ln\mu_\alpha\]\[(1 - \alpha_\alpha)K_{\bar{n}} -
\pp_{\bar{n}}\ln\mu_\alpha\]\) R_{b\bar{c}}
 \right. \nonumber \\
& & \left. + \[(1 - \alpha_\alpha)K_{\bar{n}} -
\pp_{\bar{n}}\ln\mu_\alpha\]D_b R_{m\bar{c}} + \[(1 -
\alpha_\alpha)K_m -
\pp_m\ln\mu_\alpha\]D_{\bar{n}}R_{b\bar{c}}\rbr .
\label{m5} \end{eqnarray}

This is the principal result of this section, and we will spend
the rest of the section investigating its consequences in a number
of simple examples. But first it is instructive to compare the
expression in~(\ref{m5}) with the analogous expression for the
scalar mass in equation~(4) of~\cite{Choi:1997de}.\footnote{The
analogous expression is equation~(6) of the preprint version {\tt
hep-ph/9709250} of this paper.} The two expressions share the same
general structure, though the coefficients of the various
correction terms differ -- presumably since the starting point
in~\cite{Choi:1997de} was not yet fully supersymmetric. We have
checked that~(\ref{m5}) has the proper symmetry under interchange
of indices implied by its origin from~(\ref{dddelk}).

A more substantive comparison can be made by considering a
particular K\"ahler potential. Take the case of
\begin{equation}
K = g(M,\oline{M}) + \sum_a f_a(M,\oline{M})|\Phi^a|^2 +
\frac{1}{4}\sum_{a b}X_{a b}f_a f_b|\Phi^a|^2|\Phi^b|^2 +
O(|\Phi|^3) , \label{Kquad} \end{equation}
\begin{equation} g(M,\oline{M}) = -\sum_i\ln(T^i +
\oline{T}^i),\qquad f_a(M,\oline{M}) = \prod_i(T^i +
\oline{T}^i)^{-q^a_i}, \label{orb} \end{equation}
which is motivated by modular-invariant effective actions
describing the weakly-coupled heterotic string. Then the relevant
tensors for the computation of~(\ref{m5}) are given by
\begin{eqnarray}
R_{b\bar{c}} &=& \delta_{b c}f_b\(\sum_a X_{a b} + \sum_i q_i^b\),
\qquad D_{\bar{n}}R_{b\bar{c}} = 0, \qquad D_{\bar{n}}D_b
R_{m\bar{c}} = \frac{\delta_{m n}}{2\re \;t^m}D_b R_{m\bar{c}},
\nonumber \\
D_b R_{m\bar{c}} &=& -\delta_{b c}\frac{f_b}{2\re\;
t^m}\(\sum_a\(q^b_m + q^a_m\)X_{a b} + q^b_m\sum_i q_i^b\).
\label{Rs} \end{eqnarray}
First consider the case where $F^i=0$ for the K\"ahler moduli
$T^i$ and where $F^s\oline{F}^{\bar{s}}K_{s\bar{s}} = F^s
\oline{F}^{\bar{s}}K_s K_{\bar{s}} = 3\mG^2$. Then assuming
$\pp_s\mu_\alpha = 0$ we obtain
\begin{equation} (m^2_{\Phi^I_\alpha\Pi^I_\alpha})^a_b =
3\mG^2\beta^2_\alpha\Lambda^2_\alpha \delta^a_b \(\sum_c X_{b c} +
\sum_i q_i^b\)\(1 - \alpha_\alpha\)\(2 - \alpha_\alpha\).
\label{trouble} \end{equation}
As was pointed out in~\cite{Choi:1997de}, where this same K\"ahler
potential~(\ref{Kquad}) was considered, even in this particularly
simple case of dilaton-domination (in which tree-level scalar masses
are universal and diagonal) there is a potential for sizable FCNCs
since the summation in the first term of~(\ref{trouble}) runs over
all fields which participate in the quartic coupling
of~(\ref{Kquad}). What was not appreciated in~\cite{Choi:1997de} was
the fact that the presence of this off-diagonal scalar mass
contribution depends on the parameters $\alpha_{\alpha}$, which are
determined by Planck-scale physics. In particular, the contribution
vanishes completely -- independent of the values of the modular
weights or the values of $X_{ab}$ -- provided $\alpha =1 \; {\rm or}
\; 2$.

For the untwisted sector $\Phi^a_i$, with K\"ahler potential given
by~(\ref{nskp})
\begin{equation} K^{U_i} = g_i- \ln\(1 -
e^{g_i}\sum_a|\Phi^a_i|^2\), \qquad g_i = -\ln\(T^i +
\oline{T}^i\),\end{equation}
the expression for $R_{b\bar{c}}$ is given by~(\ref{runtw}). If
the twisted sector K\"ahler potential is just the quadratic term
in~(\ref{orb}) then the twisted masses are different from the
untwisted ones. If instead the twisted sector has
\begin{equation} K^T = \sum_a e^{\sum_i  q^a_i
K^{U_i}}|\Phi^a_T|^2,\end{equation}
we would instead obtain
\begin{equation} \sum_a X_{a, b T} = \sum_i q^a_i N_i,\qquad
 \sum_a X_{a, b i} = 2N_i + \sum_{a \in T} q^a_i. \label{Xsums} \end{equation}
In orbifold constructions it is common for the quark doublet $Q_L$
of the Standard Model to arise in the untwisted sector, though
generally at least some of the other Standard Model fields must
arise in one or more of the various twisted
sectors~\cite{Giedt:2001zw,Font:1989aj}. Thus we might expect some
contributions such as those in~(\ref{Xsums}) to arise.
Nevertheless, such terms may be innocuous (quite apart from the
issue of the factors of $\alpha$) even in cases where $F^i \neq 0$
provided all twisted sector SM fields have the same modular
weights -- a condition that obtains quite often in these models.

Additional terms in~(\ref{Rs}) can arise if the K\"ahler potential
includes terms in addition to those in~(\ref{Kquad}). For example,
we might consider an addition of the form
\begin{equation}
\Delta K = \lbr \frac{1}{2} Z_{ab}\Phi^a \Phi^b +
\frac{1}{2}H_{ab\bar{c}}\Phi^a\Phi^b\oline{\Phi}^{\bar{c}} +
\frac{1}{3}Z_{abc}\Phi^a\Phi^b\Phi^c + \dots \rbr + \hc .
\label{mixholo} \end{equation}
In fact, such extensions are not uncommon in phenomenological
considerations: terms with $Z_{ab} \neq 0$ might be utilized to
generate a $\mu$-term via the Giudice-Masiero
mechanism~\cite{Giudice:1988yz} while three-field terms with both
$H_{ab\bar{c}}$ and $Z_{abc}$ non-vanishing were utilized
in~\cite{Abel:2004tt} to produce neutrino masses.

In general these terms are constrained by requirements of gauge
invariance, which do not affect the case of $X_{ab}$
in~(\ref{Kquad}). For example, a Giudice-Masiero term of $\Delta K
= \frac{1}{2}Z(T,\oline{T})H_u H_d + \hc$ can only produce
corrections to the mass matrix for the Higgs fields themselves,
and there only to the diagonal entries provided the quadratic term
in~(\ref{Kquad}) is diagonal in these fields to begin with. This
is a general property of additions to the K\"ahler potential with
the form $\Delta K = f(\Phi) + \bar{f}(\oline{\Phi})$.

More dangerous are terms with mixed holomorphicity, such as the
case of~\cite{Abel:2004tt} in which the following was added to the
K\"ahler potential
\begin{equation}
\Delta K = (Z_1)_{ab} L_a H_u N_b + (Z_2)_{ab} L_a H_d^{*} N_b +
\hc ,
\label{Abel} \end{equation}
where $N$ is a right-handed neutrino superfield, $L$ is the
standard lepton doublet of the MSSM and $a$, $b$ are generation
labels. Here we do expect contributions to off-diagonal scalar
masses (as well as additional contributions to the diagonal
entries) at one loop. For example, we have
\begin{eqnarray}
R_{L_a L_b^{*}} &=& \sum_i q_b^i f_{L_b} \delta_{ab} - \sum_c
\frac{1}{f_{N_c}}\[(Z_2)_{ac}(Z_2^*)_{bc}\frac{1}{f_{H_d}} +
(Z_1)_{ac}(Z_1^*)_{bc}\frac{1}{f_{H_u}}\] \nonumber \\
R_{N_a N_b^*} &=& \sum_i q_b^i f_{N_b} \delta_{ab} - \sum_c
\frac{1}{f_{L_c}}\[(Z_2)_{ca}(Z_2^*)_{cb}\frac{1}{f_{H_d}} +
(Z_1)_{ca}(Z_1^*)_{cb}\frac{1}{f_{H_u}}\] .
\label{lepton} \end{eqnarray}
If the coefficients $Z_1$ and $Z_2$ depend on various moduli (which
they will, in general, in realistic string models) then there are
additional terms coming from $D_{L_a}R_{T_m L_b^*}$ and
$D_{N_a}R_{T_m N_b^*}$. These terms involve derivatives of the
coefficients $Z_1$ and $Z_2$ and can introduce additional
off-diagonal terms if the $F$-terms for the corresponding moduli
fields do not vanish in the vacuum.

Though these terms must necessarily be present in this scenario,
they need {\em not} necessarily be large. The summations
in~(\ref{lepton}) are only over the various species of right-handed
neutrino and lepton doublet, respectively, that couple through the
terms in~(\ref{Abel}). While it is not inconceivable that the number
of such fields might be greater than three in string-derived
models~\cite{Giedt:2005vx,Langacker:2005pf}, it is unlikely that
such sums will generate numbers of~$\mathcal{O}(100)$. Nevertheless,
the severe constraint on the branching ratio for the rare decay $\mu
\to e \gamma$ of ${\rm Br}(\mu \to e \gamma) < 1.2 \times
10^{-11}$~\cite{Brooks:1999pu,Lavignac:2003tk} suggests that the
size of the element $m^2_{L_1 L_2^*}$ should be quite small. In
particular~\cite{Chankowski:2005jh,Gabbiani:1996hi,Feng:2000ci}
\begin{equation}
\delta^{LL}_{12} = \frac{m^2_{L_1 L_2^*}}{m^2_{L_1 L_1^*}} \lappeq
4 \times 10^{-3} ; \quad \quad {\rm for}\; \; \sqrt{m^2_{L_1
L_1^*}} = 100 \GeV,
\label{limit} \end{equation}
though the size of the off-diagonal elements can be comparable to
the diagonal entries once the typical slepton mass nears 1~TeV.
For the case of~(\ref{limit}) above, insertion of~(\ref{lepton})
into~(\ref{m5}) and summing as in~(\ref{tildek}) gives an
off-diagonal mass contribution of roughly
\begin{equation}
\delta^{LL}_{12} \simeq \frac{3N_{\nu_R}}{32\pi^2}\(|Z_1|^2 +
|Z_2^2|\)\sum_{\PV}\beta_{\PV}(1-\alpha_{\PV})(2-\alpha_{\PV})\(\frac{m_{\PV}}{m_{\PL}}\)^2
,
\end{equation}
where we have set the K\"ahler moduli to their self-dual value of
$\lang t^i \rang = 1$ in reduced Planck-mass units and taken a
dilaton-dominated scenario in which
$F^s\oline{F}^{\bar{s}}K_{s\bar{s}} = F^s \oline{F}^{\bar{s}}K_s
K_{\bar{s}} = 3\mG^2$ such that the diagonal entries of the lepton
doublet scalar mass matrix are given by $m_{3/2}^2$ (see Appendix~A
below for a complete expression). Even in the case where
$\alpha_{\PV} \neq 1,2$, and the summation is over PV fields whose
masses are very near the string scale, it is still unlikely that the
quadratically divergent contribution to FCNC in the lepton sector
would be observable except in the case of very light scalar leptons
or very large numbers of right-handed neutrino fields. In more
complicated supersymmetry-breaking scenarios the exact nature of the
FCNC bound is a more involved, model-dependent issue.
Phenomenological models which introduce higher-order terms into the
K\"ahler potential in the spirit
of~(\ref{mixholo})~\cite{Abel:2004tt,Casas:2002sn,Casas:2003kh,March-Russell:2004uf}
should consider such loop-induced contributions to FCNC processes.

\section*{Conclusion}

In this work we have considered those corrections to the scalar
potential which are quadratically dependent on the cut-off scale. We
have taken care to regulate theses divergences with a regulator
sector that preserves manifest supersymmetry. We have argued that
this step, overlooked in the past, is critical to a reliable
discussion of the physical implications of these corrections.

Specifically, these quadratically divergent contributions to the
effective scalar potential have two immediate impacts. The first
is in the determination of the vacuum expectation values of
various scalar fields in the low-energy four-dimensional theory.
Of most importance are those fields whose auxiliary fields acquire
supersymmetry breaking expectation values, as the vacuum values of
these fields will generally dominate the vacuum expectation value
of the potential itself; {\em i.e.} the vacuum energy of the
cosmos. We have shown that the sign of the loop-induced
contribution to this vacuum energy is not unambiguously positive
and thus cannot be relied upon to remedy moduli stabilization
mechanisms that produce substantial negative vacuum energy, such
as the so-called ``racetrack'' method. In fact, the sign and
magnitude of this contribution is model-dependent, but the casting
of the problem in terms of a renormalization of the spacetime
metric and K\"ahler potential -- possible only when a manifestly
supersymmetric regulating scheme is employed -- makes this
model-dependence easy to extract. We demonstrated this for the
case of dilaton stabilization in the BGW model, for which the loop
corrections were shown to affect the resulting minimum in a
negligible way.

The second, and potentially more damaging, impact of these terms is
on the generation of new contributions to the soft
supersymmetry-breaking scalar mass matrices. In this case the
constraints on the size of the off-diagonal entries of these
matrices from rare flavor-changing neutral current (FCNC) processes
makes these new contributions large enough to be worrisome if a
summation over chiral fields can be sufficient to overcome the loop
suppression factor. Of course, string theory (or any theory of
high-scale physics) must still meet the challenge of explaining the
smallness of scalar-mediated FCNC effects at the tree-level, but as
string theory is as yet unable to address the K\"ahler potential for
matter fields beyond the leading order we must ask instead for the
ways in which supergravity can spoil such a safe arrangement once it
is engineered.

When viewing this particular manifestation of the supersymmetric
flavor problem from a ground-up, phenomenological point of view it
has been common to turn away from models in which supergravity
plays a relevant role in low-energy physics so as to mitigate
these new corrections. For example, gauge meditation of
supersymmetry breaking is often promoted on this basis, and it
purports to address the problem from two fronts: (1) by making the
tree-level soft masses of the scalars proportional to gauge
charges and (2) by allowing for a drastic reduction in the size of
the gravitino mass -- and thus in the size of supergravity induced
soft-mass corrections of the form of~(\ref{bad}).

For those who take a top-down, string theory motivated point of view
the situation is less dire. String-based models are seldom
``generic'' and rarely give rise to the most general possible
supergravity effective theory. The possible K\"ahler manifolds for
the light scalars are typically quite limited form. In particular,
isometries such as $SL(2,Z)$ symmetries amongst the various moduli
(a specific form of the general K\"ahler transformation symmetry of
supergravity Lagrangians) restrict the form of the tensors appearing
in the expressions for off-diagonal scalar masses. Dangerously large
terms which mix generations are likely to result only when
particular higher-order terms are present or when particular moduli
are involved in supersymmetry breaking. Specifically, terms of three
or more fields with mixed holomorphicity tend to be the only ones of
concern. Yet large coefficients are unlikely due to gauge-invariance
constraints, except for a potential quartic term in the K\"ahler
potential. We point out that even such quartic terms tend to count
only those scalar fields within a particular sector of the string
Hilbert space as opposed to all scalar fields in the theory -- a
difference that can reduce the possible size of these corrections
substantially and which can only be appreciated by study of such
terms within the context of realistic string constructions.

It also crucially important to consider the impact of the theory
which serves as the ultraviolet (UV) completion for the supergravity
effective theory. As we demonstrate, employing a supersymmetric
regularization scheme introduces factors which must be treated as
parameters in the effective theory. These represent the
uncertainties in the threshold scale at which the UV physics (here
represented by the Pauli-Villars sector) begins to operate to
regularize divergences. The overall size of any supergravity-induced
correction to scalar mass matrices ultimately depends on these
factors in a model-dependent manner.

Even more crucial are the transformation properties of the
regularization sector under the set of gauged $U(1)_R$ symmetries
that are realized by K\"ahler transformations. The ``charges'' of
the regulating sector fields under these transformations are
determined by the dependence of the kinetic terms on the K\"ahler
potential -- parameters we denoted by $\alpha$ in~(\ref{kapPi}).
Different assumptions about this dependence act like different
regularization schemes. Some choices can make dangerous off-diagonal
scalar masses vanish identically. Such a dependence is missed by
treatments that use supersymmetry-breaking (and modular
symmetry-breaking) straight cut-off parameters and ignore the field
dependence of these cut-offs.

Can we dispense with this intrusion of Planck-scale physics upon the
quantities that interest us as low-energy observers? It is important
to recognize that the UV-dependence is inherently necessary for the
consistency of the theory. Consider, for example, the case of
anomalous $U(1)_X$ symmetries in flat (rigid) supersymmetry. In the
presence of such an anomalous $U(1)_X$ with non-vanishing $\Tr \,
T_X$ there arises a quadratically divergent contribution at one loop
proportional to $\Tr \, T_X \Lambda^2$. This contribution can not be
cancelled by $U(1)_X$-invariant PV mass terms; that is, the mass
terms in~(\ref{pvpot}) must involve fields whose $U(1)_X$ charges do
not cancel in order to generate a contribution to the quadratic
divergence. Since the kinetic terms for these fields are given by
$\exp(q_X V_X)|\Phi|^2$, where $V_X$ is the $U(1)_X$ real vector
field, their masses in~(\ref{PVfullmass}) are necessarily of the
form $M = \exp(-q_X V_X)\mu$ and are field-dependent. Anomaly
cancellation thus imposes constraints on the charges $q_X$.

The situation with K\"ahler transformations is completely analogous,
as these transformations also involve a form of $U(1)$
transformation that is anomalous at the quantum level. In the
presence of this K\"ahler anomaly there is a term
\begin{equation}
\mathcal{L}_{\rm one-loop} \ni \alpha K_{i\bar{m}} \mathcal{D}_{\mu}
z^i \mathcal{D}^{\mu} \bar{z}^{\bar{m}} \Lambda^2
\end{equation}
which can only be cancelled if there is at least some sector of the
PV fields for which $M_{\PV}^2 \propto e^{\alpha K}$. Considering
the mass term of~(\ref{PVfullmass}) we see that the PV masses can
indeed be made invariant under K\"ahler transformations if $\alpha_A
+ \alpha_B = 1$ for pairs of fields $\Phi^A$, $\Pi^B$ appearing in
the superpotential terms in~(\ref{pvpot}). Indeed, this choice gives
rise to precisely the relationship in~(\ref{PVscenA}).\footnote{This
is the configuration dubbed ``PV Scenario~A'' in the
phenomenological treatment of~\cite{Binetruy:2000md}.} This cannot
be true of {\em all} fields in the PV sector, however, since if so
the contribution of the PV sector to $\Tr \, H^2$ is always
proportional to $\frac{r}{2} + K_{i\bar{m}} \mathcal{D}_{\mu} z^i
\mathcal{D}^{\mu} \bar{z}^{\bar{m}} - 2V$ which can be removed by a
Weyl redefinition. That is, this expression vanishes identically
on-shell when the graviton equations of motion are employed.
Therefore more general choices of the various $\alpha$ values are
needed to cancel these terms for the full regularization of all
divergences (which imposes certain constraints as derived in
Section~\ref{sec:PV}) and to insure that the K\"ahler chiral anomaly
has a trace anomaly superpartner, as in the $U(1)_X$ case described
above.

In both of these examples the elements of the one-loop Lagrangian
that cannot be cancelled without field-dependent masses of the form
$m \sim e^{p\,V}$, where $p$ is a generalized ``weight'' and $V$ is
a real superfield, are elements that arise via anomalies. Therefore
we have no reason to expect that the low-energy physics should be
independent of these weights $p$: it is, after all, precisely this
property of anomalies that make them so useful in connecting UV and
IR physics. The results presented in this work are thus far more
than an academic exercise about finding the appropriate coefficients
for a complicated loop calculation, but are at the heart of how
supergravity effective Lagrangians connect low-scale physics and
high-energy assumptions in phenomenologically meaningful ways.
%

\section*{Acknowledgements}

We would like to thank G.~Kane, J.~Giedt and P.~Ko for encouragement
and helpful discussions in early stages of this work. MKG was
supported in part by the Department of Energy under grant
DE-AC02-05CH11231 and in part by the National Science Foundation
under grant PHY-0098840. BDN was supported by the Department of
Energy under grant EY-76-02-3071.

\section*{Appendix~A -- General Scalar Mass Formulae}

For the sake of completeness, in this appendix we present the
complete scalar mass expression obtained from inserting the
appropriate matrix elements into expression~(\ref{m2}) and inserting
the result into the corrected weighted sum. To keep the expression
tractable we work only to first order in the loop expansion
parameter and assume vanishing tree-level vacuum energy (i.e.
\mbox{$K_{n\bar{m}}F^n\oline{F}^{\bar{m}} = 3\mG^2$}). Then the
scalar squared mass is given by
\begin{eqnarray}
(m^2)_b^a &=& \mG^2\[\delta^a_b -
\frac{3}{32\pi^2}\(\delta^a_b\sum_p\zeta_P\eta_p\talph^2_p
\ln(\beta_p^2) \beta^2_p\Lambda^2_p - K^{a\bar{c}}
R_{b\bar{c}}\sum_\alpha\eta_\alpha
\talph_\alpha\ln(\beta_\alpha^2)\beta^2_\alpha\Lambda^2_\alpha\)\]
\nonumber \\
& & + R^a_{n\m b}F^n\oline{F}^{\m}\(1 + \frac{1}{2}\delta K -
\frac{3}{32\pi^2}\sum_p\zeta_P\eta_p\talph_p
\ln(\beta_p^2)\beta^2_p\Lambda^2_p\) \nonumber \\
& & - \loopp\sum_p\zeta_P\eta_p\[\(\bar v^p_{\m}\oline{F}^{\m} -
\mG\)v_p^{\bar l} F^n R^a_{n\bar l
b}+\hc\]\ln(\beta_p^2)\beta^2_p\Lambda^2_p \nonumber \\
& & + \loopp F^n\oline{F}^{\m}\(K^{a\bar c} R_{b\bar c}R^d_{n\m b}
+ R_{n\bar l}R^{a\;\;\bar l}_{\;\m\;\;b} +
R_{l\m}R^{a\;\;l}_{\;n\;\;b}\)
\sum_\alpha\eta_\alpha\ln(\beta_\alpha^2)\beta^2_\alpha\Lambda^2_\alpha
\nonumber \\
& & + \loopp\[ {F^n\oline{F}^{\m}\over 4x^2} \(f_n\bar f_{\bar
l}R^{a\;\;\bar l}_{\;\m\;\;b} + f_l\bar
f_{\m}R^{a\;\;l}_{\;n\;\;b}\) - m^2_{1/2}\]
\sum_\alpha\eta^\varphi_\alpha\ln(\beta^\varphi_\alpha)^2
(\beta^\varphi_\alpha)^2\Lambda_\varphi^2 \nonumber \\
& & - \loopp F^n\oline{F}^{\m}K^{a\bar c}\sum_\alpha\(D_{\m}D_b
R_{n\bar c} + \bar v^\alpha_{\m} D_b R_{n\bar c} + v^\alpha_n
D_{\m}R_{b\bar c} + v^\alpha_n\bar v^\alpha_{\m}R_{b\bar c}\)
\eta_\alpha\ln(\beta_\alpha^2)\beta^2_\alpha\Lambda^2_\alpha
\nonumber \\
& & - {\delta^a_b\over16\pi^2}F^n
\oline{F}^{\m}R_{n\m}\sum_\alpha\eta_\alpha\talph_\alpha
\ln(\beta_\alpha^2)\beta^2_\alpha\Lambda^2_\alpha -
\loopp\sum_p\zeta_P\eta_p\alpha_p|v_n^p
F^n|^2\ln(\beta_p^2)\beta^2_p\Lambda^2_p, \end{eqnarray}
where
\begin{eqnarray}
\talph_p &=& (\talph_\alpha,\wh{\talph}_\alpha,\talph^\phi_\alpha,
\talph^\theta_b) = (1 - \alpha_\alpha,-\hat\alpha_\alpha,0,
\alpha^\theta_b), \nonumber \\
v^\alpha_i &=& \talph_\alpha K_i - \pp_i\mu - \Gamma^k_{k i},
\qquad (v^\varphi_\alpha)_i = K_i - {f_i\over2x} -
\pp_i\mu_\varphi, \nonumber \\
v^p_i &=& \talph_p K_i - \pp_i\mu_P,\qquad P_p =
\wh\Phi,\wh\Pi,\theta_b,\qquad v_p^{\m} = K^{\m
i}v^p_i.\label{talph}  \end{eqnarray}
If we take the specific case of K\"ahler potential given
by~(\ref{Kquad})
\begin{equation}
K = g(M,\oline{M}) + \sum_a f_a(M,\oline{M})|\Phi^a|^2 +
\frac{1}{4}\sum_{a b}X_{a b}f_a f_b|\Phi^a|^2|\Phi^b|^2 +
O(|\Phi|^3) , \end{equation}
which was also the example considered in~\cite{Choi:1997de}, we
obtain for the diagonal scalar mass entries
\begin{eqnarray}
m^2_a &=& \mG^2\[1 - \frac{3}{32\pi^2} \(\sum_p \zeta_P \eta_p
\talph^2_p \ln(\beta_p^2) \beta^2_p\Lambda^2_p - \(\sum_i q^a_i +
\sum_b X_{a b}\)\sum_\alpha \eta_\alpha \talph_\alpha
\ln(\beta_\alpha^2) \beta^2_\alpha\Lambda^2_\alpha\)\] \nonumber
\\
& & - q^a_i g_{i\jbar}\; F^i\oline{F}^{\jbar} \(1 +
\frac{1}{2}\delta K - \frac{3}{32\pi^2} \sum_p \zeta_P \eta_p
\talph_p \ln(\beta_p^2)\beta^2_p\Lambda^2_p\) \nonumber \\
 & & + q^a_i \loopp \sum_p \zeta_P \eta_p
 \[\(\bar v^p_{\m} \oline{F}^{\m} -
\mG\) v_i^p F^i+\hc\] \ln(\beta_p^2)\beta^2_p\Lambda^2_p \nonumber
\\
& & - \loopp g_{i\jbar}\; F^i\oline{F}^{\jbar}\[2\talph\(2 +
\sum_b q^b_i\) - \sum_b q^b_i X_{a b} \right. \nonumber \\
& & \left. \qquad \qquad - 2q^a_i\(2 + \sum_b X_{a b} +\sum_k
q^a_k + \sum_b q^b_i\) \]
\sum_\alpha\eta_\alpha\ln(\beta_\alpha^2)\beta^2_\alpha\Lambda^2_\alpha
\nonumber \\
& & - \loopp\[ m^2_{1/2} \sum_\alpha \eta^\varphi_\alpha
\ln(\beta^\varphi_\alpha)^2
(\beta^\varphi_\alpha)^2\Lambda_\varphi^2 + \sum_i
q^a_i\sum_\alpha|v^\alpha_n F^n|^2
\eta_\alpha\ln(\beta_\alpha^2)\beta^2_\alpha|\Lambda^2_\alpha\]
\nonumber \\
 & & - \loopp\[\sum_b\(q^a_i + q^b_i\)X_{a b} +
q^b_i \sum_k q^a_k\] \(g_i F^i\oline{F}^{\m}\sum_\alpha\bar
v^\alpha_{\m} + \hc\) \eta_\alpha \ln(\beta_\alpha^2)
\beta^2_\alpha|\Lambda^2_\alpha \nonumber \\
& &  - {\delta^a_b\over16\pi^2}F^s\oline{F}^{\s}R_{s\s}
\sum_\alpha\eta_\alpha\talph_\alpha
\ln(\beta_\alpha^2)\beta^2_\alpha\Lambda^2_\alpha -
\loopp\sum_p\zeta_P\eta_p\alpha_p|v_n^p
F^n|^2\ln(\beta_p^2)\beta^2_p\Lambda^2_p, \end{eqnarray}
where $n,m = s,i,j,k$ and $i,j,k$ label T-moduli.

\section*{Appendix~B -- Errata}

In this appendix we take a moment to correct a handful of errata
in some of the papers referred to in this work:
\begin{description}
\item [Ref.~\cite{Gaillard:1994sf}] In the first line of Eq.~(7)
the term $-\frac{r}{2}$ should be replaced by $+\frac{7r}{2}$.
This entails a correction to Eqs.~(15), (16) and~(20) of that
paper; the correct relations are given
in~\cite{Gaillard:1998bf,Gaillard:1999ir}. In Eq.~(17) there
should be a replacement $\wh{V} \to V$ in the first term on the
right-hand side. In this same equation, the following term
\begin{equation} \half H_b^2\ni {4\over x
x_b}|\delta^b\theta^b|^2K_{b\bar b}\D_a D_b(T^a z)^b = -{1\over
x}\delta k_b\Gamma^b_{b k}(T^a z)^k\D_a =
-2\alpha_b^\theta\D\delta k_b . \label{missing}\end{equation}
is missing and should be inserted. This term can be spotted in the
4th term on the right-hand side of Eq.~(C.36)
in~\cite{Gaillard:1996ms}; it corresponds to the last term in
Eq.~(C.37) of the same reference.
\item [Ref.~\cite{Gaillard:1996hs}] In the expression for $L_0$ in
Eq.~(13), the coefficient of $\frac{r}{2}$ in the first term
should be $(N_G + 7)$ instead of $(N_G - 1)$ and the second term
should have $2K_{i\bar m}\({\cal D}_{\mu} z^i {\cal D}^{\mu}
\bar{z}^{\bar{m}} + 5F^i \oline{F}^{\bar{m}}\)$ replaced by
$2K_{i\bar{m}}\(2{\cal D}_{\mu} z^i {\cal D}^{\mu}
\bar{z}^{\bar{m}} - 5F^i \oline{F}^{\bar{m}}\)$.
\end{description}
%


\begin{thebibliography}{99}
\bibitem{Ellis:1981ts}
  J.~R.~Ellis and D.~V.~Nanopoulos,
  Phys.\ Lett.\ B {\bf 110}, 44 (1982).
\bibitem{Donoghue:1983mx}
  J.~F.~Donoghue, H.~P.~Nilles and D.~Wyler,
  Phys.\ Lett.\ B {\bf 128}, 55 (1983).
\bibitem{Louis:1994ht}
  J.~Louis and Y.~Nir,
  Nucl.\ Phys.\ B {\bf 447}, 18 (1995).
\bibitem{Chankowski:2005jh}
  P.~H.~Chankowski, O.~Lebedev and S.~Pokorski,
  Nucl.\ Phys.\ B {\bf 717}, 190 (2005).
\bibitem{Lebedev:2005uh}
  O.~Lebedev,
  ``A stringy solution to the FCNC problem,''
  [hep-ph/0506052].
\bibitem{Barbieri:1982eh}
  R.~Barbieri, S.~Ferrara and C.~A.~Savoy,
  Phys.\ Lett.\ B {\bf 119}, 343 (1982).
\bibitem{Chamseddine:1982jx}
  A.~H.~Chamseddine, R.~Arnowitt and P.~Nath,
  Phys.\ Rev.\ Lett.\  {\bf 49}, 970 (1982).
\bibitem{Binetruy:2005ez}
  P.~Binetruy, G.~L.~Kane, J.~D.~Lykken and B.~D.~Nelson,
  ``Twenty-five questions for string theorists,'' (2005)
  [arXiv:hep-th/0509157].
\bibitem{Choi:1994xg}
  K.~Choi, J.~E.~Kim and H.~P.~Nilles,
  Phys.\ Rev.\ Lett.\  {\bf 73}, 1758 (1994).
\bibitem{Choi:1997de}
  K.~Choi, J.~S.~Lee and C.~Munoz,
  Phys.\ Rev.\ Lett.\  {\bf 80}, 3686 (1998).
%
%
\bibitem{Coleman:1973jx}
  S.~R.~Coleman and E.~Weinberg,
  Phys.\ Rev.\ D {\bf 7}, 1888 (1973).
\bibitem{Srednicki:1984un}
  M.~Srednicki and S.~Theisen,
  Phys.\ Rev.\ Lett.\  {\bf 54}, 278 (1985).
\bibitem{Breit:1985ns}
  J.~D.~Breit, B.~A.~Ovrut and G.~Segre,
  Phys.\ Lett.\ B {\bf 162}, 303 (1985).
\bibitem{Binetruy:1985ap}
  P.~Binetruy and M.~K.~Gaillard,
  Phys.\ Lett.\ B {\bf 168}, 347 (1986).
\bibitem{Chiou-Lahanas:1989sf}
  C.~Chiou-Lahanas, A.~Kapella-Economu, A.~B.~Lahanas and X.~N.~Maintas,
  Phys.\ Rev.\ D {\bf 42}, 469 (1990).
\bibitem{Chiou-Lahanas:1992am}
  C.~Chiou-Lahanas, A.~Kapella-Economou, A.~B.~Lahanas and X.~N.~Maintas,
  Phys.\ Rev.\ D {\bf 45}, 534 (1992).
\bibitem{Gaillard:1993es}
  M.~K.~Gaillard and V.~Jain,
  Phys.\ Rev.\ D {\bf 49}, 1951 (1994).
\bibitem{Gaillard:1996hs}
  M.~K.~Gaillard, V.~Jain and K.~Saririan,
  Phys.\ Lett.\ B {\bf 387}, 520 (1996).
\bibitem{Gaillard:1996ms}
  M.~K.~Gaillard, V.~Jain and K.~Saririan,
  Phys.\ Rev.\ D {\bf 55}, 883 (1997).
\bibitem{Binetruy:1987cn}
  P.~Binetruy, S.~Dawson, M.~K.~Gaillard and I.~Hinchliffe,
  Phys.\ Rev.\ D {\bf 37}, 2633 (1988).
\bibitem{Burton:1989ai}
  J.~W.~Burton, M.~K.~Gaillard and V.~Jain,
  Phys.\ Rev.\ D {\bf 41}, 3118 (1990).
\bibitem{Giedt:2001zw}
  J.~Giedt,
  Annals Phys.\  {\bf 297}, 67 (2002).
\bibitem{Binetruy:1988dw}
  P.~Binetruy and M.~K.~Gaillard,
  Phys.\ Lett.\ B {\bf 220}, 68 (1989).
\bibitem{Binetruy:1990hf}
  P.~Binetruy and M.~K.~Gaillard,
  Nucl.\ Phys.\ B {\bf 358}, 121 (1991).
\bibitem{Gaillard:1992bt}
  M.~K.~Gaillard and T.~R.~Taylor,
  Nucl.\ Phys.\ B {\bf 381}, 577 (1992).
\bibitem{Gaillard:1994sf}
  M.~K.~Gaillard,
  Phys.\ Lett.\ B {\bf 342}, 125 (1995).
\bibitem{bgpr} A.~Birkedal, M.~K.~Gaillard,
  C.~Park and M.~Ransdorp, paper in preparation.
\bibitem{Binetruy:1988nx}
  P.~Binetruy and M.~K.~Gaillard,
  Nucl.\ Phys.\ B {\bf 312}, 341 (1989).
\bibitem{Antoniadis:1991fh}
  I.~Antoniadis, K.~S.~Narain and T.~R.~Taylor,
  Phys.\ Lett.\ B {\bf 267}, 37 (1991).
\bibitem{Dixon:1990pc}
  L.~J.~Dixon, V.~Kaplunovsky and J.~Louis,
  Nucl.\ Phys.\ B {\bf 355}, 649 (1991).
\bibitem{Choi:1997cm}
  K.~Choi, H.~B.~Kim and C.~Munoz,
  Phys.\ Rev.\ D {\bf 57}, 7521 (1998).
%
%
\bibitem{Gaillard:1998bf}
  M.~K.~Gaillard,
  Phys.\ Rev.\ D {\bf 58}, 105027 (1998).
\bibitem{Gaillard:1999ir}
  M.~K.~Gaillard,
  Phys.\ Rev.\ D {\bf 61}, 084028 (2000).
\bibitem{Gaillard:2000fk}
  M.~K.~Gaillard and B.~D.~Nelson,
  Nucl.\ Phys.\ B {\bf 588}, 197 (2000).
\bibitem{Binetruy:2000md}
  P.~Binetruy, M.~K.~Gaillard and B.~D.~Nelson,
  Nucl.\ Phys.\ B {\bf 604}, 32 (2001).
%
%
\bibitem{Binetruy:1996nx}
  P.~Binetruy, M.~K.~Gaillard and Y.~Y.~Wu,
  Nucl.\ Phys.\ B {\bf 493}, 27 (1997).
\bibitem{Binetruy:1997vr}
  P.~Binetruy, M.~K.~Gaillard and Y.~Y.~Wu,
  Phys.\ Lett.\ B {\bf 412}, 288 (1997).
\bibitem{Binetruy:1996xj}
  P.~Binetruy, M.~K.~Gaillard and Y.~Y.~Wu,
  Nucl.\ Phys.\ B {\bf 481}, 109 (1996).
\bibitem{Casas:1996zi}
  J.~A.~Casas,
  Phys.\ Lett.\ B {\bf 384}, 103 (1996).
\bibitem{Shenker:1990uf}
  S.~H.~Shenker,
  ``The Strength Of Nonperturbative Effects In String Theory,''
RU-90-47
{\it Presented at the Cargese Workshop on Random Surfaces, Quantum
Gravity and Strings, Cargese, France, May 28 - Jun 1, 1990}.
\bibitem{Silverstein:1996xp}
  E.~Silverstein,
  Phys.\ Lett.\ B {\bf 396}, 91 (1997).
\bibitem{Banks:1994sg}
  T.~Banks and M.~Dine,
  Phys.\ Rev.\ D {\bf 50}, 7454 (1994).
\bibitem{Giedt:2003ap}
  J.~Giedt and B.~D.~Nelson,
  JHEP {\bf 0405}, 069 (2004).
\bibitem{Gaillard:1999et}
  M.~K.~Gaillard and B.~D.~Nelson,
  Nucl.\ Phys.\ B {\bf 571}, 3 (2000).
\bibitem{Birkedal-Hansen:2001is}
  A.~Birkedal-Hansen and B.~D.~Nelson,
  Phys.\ Rev.\ D {\bf 64}, 015008 (2001).
\bibitem{Birkedal-Hansen:2002am}
  A.~Birkedal-Hansen and B.~D.~Nelson,
  Phys.\ Rev.\ D {\bf 67}, 095006 (2003).
\bibitem{Kane:2002ap}
  G.~L.~Kane, J.~D.~Lykken, B.~D.~Nelson and L.~T.~Wang,
  Phys.\ Lett.\ B {\bf 551}, 146 (2003).
\bibitem{Barreiro:1998nd}
  T.~Barreiro, B.~de Carlos, J.~A.~Casas and J.~M.~Moreno,
  Phys.\ Lett.\ B {\bf 445}, 82 (1998).
\bibitem{Gaillard:2003gt}
  M.~K.~Gaillard, J.~Giedt and A.~L.~Mints,
  Nucl.\ Phys.\ B {\bf 700}, 205 (2004)
  [Erratum-ibid.\ B {\bf 713}, 607 (2005)].
\bibitem{Witten:1996mz}
  E.~Witten,
  Nucl.\ Phys.\ B {\bf 471} (1996) 135.
%
%
\bibitem{Font:1989aj}
  A.~Font, L.~E.~Ibanez, F.~Quevedo and A.~Sierra,
  Nucl.\ Phys.\ B {\bf 331}, 421 (1990).
\bibitem{Ibanez:1992hc}
  L.~E.~Ibanez and D.~Lust,
  Nucl.\ Phys.\ B {\bf 382}, 305 (1992).
\bibitem{Kane:2002qp}
  G.~L.~Kane, J.~D.~Lykken, S.~Mrenna, B.~D.~Nelson, L.~T.~Wang and T.~T.~Wang,
  Phys.\ Rev.\ D {\bf 67}, 045008 (2003).
\bibitem{Kane:2004tk}
  G.~L.~Kane, T.~T.~Wang, B.~D.~Nelson and L.~T.~Wang,
  Phys.\ Rev.\ D {\bf 71}, 035006 (2005).
\bibitem{Giudice:1988yz}
  G.~F.~Giudice and A.~Masiero,
  Phys.\ Lett.\ B {\bf 206}, 480 (1988).
\bibitem{Abel:2004tt}
  S.~Abel, A.~Dedes and K.~Tamvakis,
  Phys.\ Rev.\ D {\bf 71}, 033003 (2005).
\bibitem{Giedt:2005vx}
  J.~Giedt, G.~L.~Kane, P.~Langacker and B.~D.~Nelson,
  Phys.\ Rev.\ D {\bf 71}, 115013 (2005).
\bibitem{Langacker:2005pf}
  P.~Langacker and B.~D.~Nelson,
  Phys.\ Rev.\ D {\bf 72}, 053013 (2005).
\bibitem{Brooks:1999pu}
  M.~L.~Brooks {\it et al.}  [MEGA Collaboration],
  Phys.\ Rev.\ Lett.\  {\bf 83}, 1521 (1999).
\bibitem{Lavignac:2003tk}
  S.~Lavignac,
  ``Flavour and CP violation in the lepton sector and new physics,''
  eConf {\bf C030603}, VEN04 (2003)
  [arXiv:hep-ph/0312309].
\bibitem{Gabbiani:1996hi}
  F.~Gabbiani, E.~Gabrielli, A.~Masiero and L.~Silvestrini,
  Nucl.\ Phys.\ B {\bf 477}, 321 (1996).
\bibitem{Feng:2000ci}
  J.~L.~Feng,
  ``Theoretical motivations for lepton flavor violation,'' (2001)
  [arXiv:hep-ph/0101122].
\bibitem{Casas:2002sn}
  J.~A.~Casas, J.~R.~Espinosa and I.~Navarro,
  Phys.\ Rev.\ Lett.\  {\bf 89}, 161801 (2002).
\bibitem{Casas:2003kh}
  J.~A.~Casas, J.~R.~Espinosa and I.~Navarro,
  JHEP {\bf 0309}, 048 (2003).
\bibitem{March-Russell:2004uf}
  J.~March-Russell and S.~M.~West,
  Phys.\ Lett.\ B {\bf 593}, 181 (2004).

\end{thebibliography}

\end{document}